\documentclass[osajnl,twocolumn,showpacs,superscriptaddress,11pt]{revtex4-1} 
\usepackage{graphicx}
\usepackage{subfig}

\usepackage{amsmath,amssymb,amsfonts}       

\usepackage{hyperref}%

\usepackage{bbm}
\usepackage{float}
\usepackage{color}

\newcommand{\TFd}{\mathop{\mathbf{TF}}}

\newcommand{\average}[1]{\left \langle {#1} \right \rangle}

\newcommand{\T}{\mathsf{T}} 
\newcommand{\0}{\mathsf{0}} 
\newcommand{\I}{\mathbf{I}} 
\newcommand{\Hkal}{\mathbf{H}_{\infty}} 
\newcommand{\Lkal}{\mathbf{L}_{\infty}} 
\newcommand{\K}{\mathbf{K}_{\infty}} 
\newcommand{\z}{{z}} 
\DeclareMathOperator{\sinc}{sinc}

\newcommand{\CovMat}{\boldsymbol{\Sigma}} 

\newcommand{\xz}{\text{\bf X}} 
\newcommand{\uz}{\text{\bf U}} 
\newcommand{\yz}{\text{\bf S}} 

\newcommand{\propend}{\hfill \ensuremath{\bigtriangleup}}

\newcommand{\dint}{\mathrm{d}} 

\newcommand{\Ad}{\mathcal{A}_\mathsf{tur}} 
\newcommand{\Bd}{\mathcal{B}_\mathsf{tur}} 
\newcommand{\Cd}{\mathcal{C}_\mathsf{tur}} 

\newcommand{\rhovec}{\boldsymbol{\rho}}
\newcommand{\kappavec}{\boldsymbol{\kappa}}
\newcommand{\etavec}{{\boldsymbol{\eta}}} 
\newcommand{\phivec}{{\boldsymbol{\psi}}} 
\newcommand{\varphivec}{{\boldsymbol{\varphi}}} 
 
\newcommand{\svec}{{\mathbf{s}}} 
\newcommand{\xvec}{{\mathbf{x}}} 
 
\newcommand{\uvec}{{\mathbf{u}}}

 \newcommand{\FTR}{\widetilde{\mathcal{R}}} 
  \newcommand{\G}{\mathcal{G}} 
  \newcommand{\FTG}{\widetilde{\mathcal{G}}} 
  \newcommand{\FTP}{\widetilde{\Pi}} 
  \newcommand{\Sl}{\mathbf{s}} 
  \newcommand{\FTS}{\widetilde{\mathbf{s}}} 
  \newcommand{\WF}{\boldsymbol{\varphi}} 
  \newcommand{\FTWF}{\widetilde{\boldsymbol{\varphi}}} 
  \newcommand{\Noise}{\boldsymbol{\eta}} 
  \newcommand{\FTN}{\widetilde{\boldsymbol{\eta}}} 

  \newcommand{\Sphi}{\mathbf{W}_\varphi} 
  
\newtheorem{theorem}{Theorem}[section]

\newtheorem{property}[theorem]{Property}

\newenvironment{example}[1][Example]{\begin{trivlist}
\item[\hskip \labelsep {\bfseries #1}]}{\end{trivlist}}

\newcommand{\qed}{\nobreak \ifvmode \relax \else
      \ifdim\lastskip<1.5em \hskip-\lastskip
      \hskip1.5em plus0em minus0.5em \fi \nobreak
      \vrule height0.75em width0.5em depth0.25em\fi}

\newcommand{\pushright}[1]{\ifmeasuring@#1\else\omit\hfill$\displaystyle#1$\fi\ignorespaces}
\newcommand{\pushleft}[1]{\ifmeasuring@#1\else\omit$\displaystyle#1$\hfill\fi\ignorespaces}

\begin{document}

\title{Modelling astronomical adaptive optics performance with temporally-filtered Wiener 
  reconstruction of slope data}
\author{Carlos M. Correia}\email{carlos.correia@lam.fr} 
\affiliation{Aix Marseille Univ, CNRS, LAM (Laboratoire d'Astrophysique de Marseille) UMR 7326, 13388, Marseille, France}
\author{Charlotte Z. Bond}
\affiliation{Aix Marseille Univ, CNRS, LAM (Laboratoire d'Astrophysique de Marseille) UMR 7326, 13388, Marseille, France}

 \author{Jean-Fran\c{c}ois Sauvage}
 \affiliation{Aix Marseille Univ, CNRS, LAM (Laboratoire
   d'Astrophysique de Marseille) UMR 7326, 13388, Marseille, France}
 \affiliation{ONERA -- Office National d'Etudes et de Recherches
   A\'erospatiales), B.P.72, F-92322 Ch\^atillon, France}

\author{Thierry Fusco}
 \affiliation{Aix Marseille Univ, CNRS, LAM (Laboratoire
   d'Astrophysique de Marseille) UMR 7326, 13388, Marseille, France}
 \affiliation{ONERA -- Office National d'Etudes et de Recherches
   A\'erospatiales), B.P.72, F-92322 Ch\^atillon, France}

\author{Rodolphe Conan}
\affiliation{Giant Magellan Telescope Corporation, 465 N. Halstead Street, Suite 250
Pasadena, CA 91107, USA}
\author{Peter Wizinowich}
\affiliation{WM Keck Observatory, 65-1120 Mamalahoa Hwy
Kamuela, HI 96743, USA}



\begin{abstract}
We build on a long-standing tradition in Astronomical Adaptive
Optics (AO) of specifying performance metrics and error budgets using linear systems 
modeling in the spatial-frequency domain. 
\vspace{5pt}

Our goal is to provide a comprehensive tool for the calculation
of error budgets in terms of residual temporally-filtered phase power-spectral-densities (PSDs)
and variances. In addition, the fast simulation of AO-corrected
PSFs provided by this method can be used as inputs for simulations of 
science observations with next-generation instruments and telescopes,
in particular to predict post-coronagraphic constrast improvement for
planet finder systems.

\vspace{5pt}
We extend the previous results presented in 
\textit{Correia et al }\cite{correia14a} to the closed-loop case with
predictive controllers and generalise the
analytical modelling of \textit{Rigaut et al }\cite{rigaut98},
\textit{Flicker et al }\cite{flicker07a} and \textit{Jolissaint} \cite{jolissaint10}. We
follow closely the developments of \textit{Ellerbroek}
\cite{ellerbroek05} and propose the synthesis of a distributed Kalman
filter to mitigate both aniso-servo-lag and aliasing errors whilst minimizing the overall residual variance. 
 We discuss
applications to (\textit{i}) Analytic AO-corrected
point-spread-function (PSF) modelling in the spatial-frequency
  domain; (\textit{ii}) Post-coronagraphic  contrast enhancement; 
  (\textit{iii}) Filter optimisation for real-time wave-front
  reconstruction; and (\textit{iv}) PSF reconstruction from system
  telemetry.

\vspace{5pt}
Under perfect knowledge of wind-velocities we show that $\sim$ 60\,nm\,rms
error reduction can be achieved with the distributed Kalman Filter
embodying anti-aliasing reconstructors on 10\,m-class high-order AO
systems, leading to contrast
improvement factors of up to three orders of magnitude at few
$\lambda/D$ separations ($\sim1-5 \lambda/D$) for a 0-magnitude star and reaching close to 1 order of
magnitude for a 12-magnitude star. 


\end{abstract}

\maketitle





  \section{Introduction}\label{sec:intro}
Synthetic modelling of Adaptive Optics (AO) systems has been pursued
over the years using linear approximations.  These approximations allow 
a fast and accurate estimation of performance over a broad parameter
range for both
classical (single-conjugate) AO~\cite{rigaut98,flicker07a,jolissaint10} and
multi-conjugate AO~\cite{rigaut00, gavel04, ellerbroek05}. 
This is particularly relevant in the era of Extremely Large Telescopes
(ELTs), where a high number of degrees of freedom incur high computational costs
in full end-to-end simulations and likewise in real-time during on-sky
observations.

In \textit{Correia et al}~\cite{correia14a}
post-facto spatial anti-aliasing filters were investigated.
Here we generalize these results to the
dynamic, temporally-filtered case, in particular for regular closed-loop
operation.  The temporal filtering of the loop is factored in by invoking, as
elsewhere in the literature, the frozen-flow hypothesis, for which
there is growing evidence\cite{ono16, poyneer09}. 

A general state-space framework is employed to synthesise the
closed-loop linear filters -- a specific case being the commonly-used
integrator controller --
borrowing from the motivation and further developing the results 
of~\cite{poyneer07, massioni11} on distributed filters. 
The latter relies on the hypothesis of phase spatial invariance. As such, the AO loop operations can all be approximated by convolutions with
localised kernels which translate ~into filtering functions when a basis
of complex exponentials is chosen.  All parameters being equal,
the two approaches above are equivalent but differ in how the controller is applied 
to the measurements.  In direct space the data is
convolved (after inverse Fourier-transforming the matrix of gains).
In Fourier space, the Fourier transform of the wave-front measurements
is taken, then each mode is weighted by a complex function
(the filtering process) followed by an inverse Fourier transform to gather the
reconstructed phase.  

Although this is not a review paper we
strive to generalise and provide further insight into previous
work in this field.
Our developments make extensive use of
Fourier transforms, in both time and space over
continuous and discrete supports (with the necessary adaptations), and
the well-established relationships between the Laplace transform 
and \emph{Z-transforms}~\cite{oppenheim97,oppenheim99}. 

The foreseeable applications of these results include
\begin{itemize}
\item Analytic AO-corrected PSF modelling 
\item Post-coronagraphic contrast enhancement
\item Filter optimisation for real-time wave-front reconstruction
\item Evaluation of error terms for post-processing system telemetry to
  reconstruct the AO-corrected PSF
\end{itemize} 

Of particular interest for the analysis
carried out here are cases where the use of an optical spatial filter \cite{poyneer04a}
is limited, such as in cases where the size of the sub-apertures compared to
the turbulence coherent length makes it hard to reach a suitable
trade-off between aliasing rejection (small spatial filter width) and
robustness to changing seeing conditions (large spatial filter
width)~\cite{sauvage16}.  In such cases a reconstruction process which
can compensate for predicted aliasing errors is highly desirable, particularly in the
case of high contrast imaging.

The formulation of the distributed Kalman filter (DKF) is motivated by physical and technological
constraints.  On the one hand complex-exponential (Fourier) modes are considered statistically
independent (both spatially and temporally).  On the other hand, if $N$ is
the number of degrees of freedom,  the runtime computation of modal
coefficients can be accomplished using transforms that scale with $N\log(N)$,
compared to $N^2$ operations required by standard linear-systems solvers
(explicit vector-matrix multiplication).  Thus such an approach has the potential 
to provide a much welcome improvement in runtime application.
Moreover, since a distributed controller is obtained, the
off-line computation of gains boils down to solving scalar (or small
scale) algebraic equations in parallel for each Fourier mode. 

Considering that Kalman filters are seamlessly obtained as the minimisers of the
residual phase variance criterion (which is to say Strehl-ratio
maximisers) they naturally achieve the best trade-off between aniso-servo-lag,
spatial aliasing and propagated noise. 
Mitigation of these errors is crucial for achieving
high post-coronagraphic contrast levels sufficiently close to the PSF core
to enable the detection of Earth-mass companions orbiting nearby
stars \cite{guyon05}. 

Illustrative examples are given for two scenarios
\begin{enumerate}
\item PSF modelling and performance optimisation for current general-purpose AO systems
\item Post-coronagraphic raw constrast
  improvement in current 10m-class
  telescopes 
\end{enumerate}
using a distributed Kalman filter
\cite{massioni11, poyneer08a} with embedded temporal prediction of
the atmosphere. 

This paper is organised in the following way: Section~\ref{sec:WF_alias_noise_error} 
formulates the residual wave-front
error statistics in the static and dynamic cases;
Sec.~\ref{sec:controller-synthesis} provides the distributed Kalman
Filter formulation and reviews that of the integrator as a specific
case covered by state-space models;
Sec.~\ref{sec:SHmodelling} outlines the modelling for Shack-Hartmann
based AO; Sec.\ref{sec:postCoronaConstrast} and
Sec.~\ref{sec:examples} quantifies AO performance and raw contrast
improvement when using optimal, time-predicive controllers; 
a summary is provided in Sec.~\ref{sec:summary}.
  \section{Residual  wave-front error in the spatial-frequency domain}\label{sec:WF_alias_noise_error}

The errors present in an adaptive optics system can be
separated into different categories according to their
origin.  Depending on the nature of the error they may be
static or impacted by the AO loop.  The specific errors
considered in this paper are:
\begin{enumerate}
	\item \emph{Fitting error} for  modes beyond the AO
          correction region, \textit{i.e. } spatial frequencies above
          $\frac{1}{2d}$, where $d$ is the deformable mirror pitch.
	\item \emph{Measurement noise} due to photon and detector read
         noise. 
	\item \emph{Aliasing} due to spurious high-order frequencies which
          fold into the measurement during the discretisation process.
        ~\cite{correia14a}.  
	\item \emph{Aniso-servo-lag errors} due to system delays
          and the angular separation between guide-stars and science objects.
\end{enumerate}


We consider two cases -- static and dynamic, i.e., with and without
temporal filtering by an AO loop, in both open- and closed-loop
mode. 

The telescope aperture, which would break spatial stationarity
is approximated by the general filtering functions for piston-removal
\cite{correia14a, ellerbroek05}. Therefore the final result is compatible with expressions in the spatial
frequency domain using modal decompositions onto functions defined
over an infinite plane. Tilde symbols are used to represent
complex-valued variables in the spatial-frequency domain, each of
which represent coefficients of harmonic functions whose relationship to
spatial domain variables are given by standard Fourier transforms \cite{oppenheim97}.

\subsection{Limiting static case: open-loop, $\tau\rightarrow0$, $T_s\rightarrow0$}
We start by providing suitable formulations for evaluating the individual
AO error contributions in the highly idealised static case where any
temporal aspects are momentarily set aside. This case is rather
instructive as the limiting case when the WFS integration time is
zero ($T_s=0$) and the delays are set to null ($\tau = 0$) leading to
simplifications when computing the instantaneous AO
residual. Moreover, no time-dependence is associated to the wave-front
sensor $\widetilde{\mathcal{G}}(\boldsymbol{\kappa})$, spatial reconstruction
$\widetilde{\mathcal{R}}(\boldsymbol{\kappa})$ and deformable
mirror $
\widetilde{\mathcal{F}}(\boldsymbol{\kappa})$ filtering
functions. Spatial filters are 
indexed by the spatial frequency variable $\boldsymbol{\kappa} =
\{\kappa_x ;\kappa_y\} \in \mathbb{R}^2$. 

\begin{figure}[htpb]
	\begin{center}
            \includegraphics[width=0.45\textwidth]{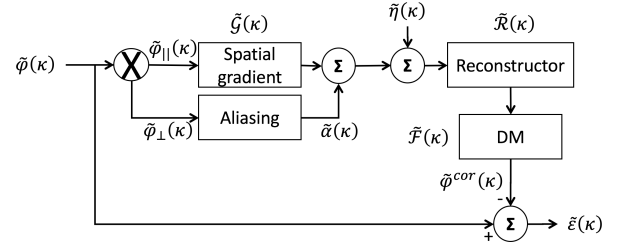}
	\end{center}
	\caption[]
	{\label{fig:loopOL}
     Block-diagram for open-loop operation.  See text for definition of
     variables.
     }
\end{figure}

Figure \ref{fig:loopOL} provides the general-case AO scenario  in
open-loop operation. Variables therein will be used throughout this
document. 

The (spatial) error functions,
$\widetilde{\varepsilon}(\boldsymbol{\kappa})$ are described by the following relations
\begin{equation}\label{eq:errFunctionsExpanded}
\left[
\begin{array}[c]{cc}%
\tilde\varepsilon_{||}(\boldsymbol{\kappa})\\\tilde\varepsilon_\perp(\boldsymbol{\kappa})
    \end{array}\right] = \left[
\begin{array}[c]{cc}%
\FTWF_{||}(\boldsymbol{\kappa})\\\FTWF_\perp(\boldsymbol{\kappa})
    \end{array}\right]
-
\left[
\begin{array}[c]{cc}%
\FTWF_{||}^{\text{cor}}(\boldsymbol{\kappa})\\\FTWF_\perp^{\text{cor}}(\boldsymbol{\kappa})
    \end{array}\right]
\end{equation}
where $||$ represents the AO controllable region $-1/(2d) \leq
\boldsymbol{\kappa} \leq 1/(2d)$ and $\perp$ its complement for a DM
with regular actuator pitch $d$ matching the SH-WFS sub-aperture widths. $\FTWF$ refers to wave-fronts in the spatial-frequency
domain.
The upper-script $'\text{cor}'$
represents the corrective phase applied by the DM. Note that in
general $\FTWF_\perp^{\text{cor}}(\boldsymbol{\kappa})=0$ but we include this term
as the shape of the DM actuator functions can extend 
the controllable region beyond the limits stated above
\cite{ellerbroek05, correia14a}. In the
following we drop the symbol $||$ when implicit and use
interchangeably the notation for the power-spectral density (PSD) $\average{|v|^2}(\boldsymbol{\kappa}) = \mathbf{W}_v(\boldsymbol{\kappa}) $.

Let the piston-removed residual phase PSD be such that
\begin{equation}\label{eq:PRfilter}
\Sphi'(\boldsymbol{\kappa}) = \left[1-\left|\frac{2 J_1(\pi \boldsymbol{\kappa} D)}{\pi \boldsymbol{\kappa} D} \right|^2 \right] \Sphi(\boldsymbol{\kappa}) =  \widetilde{\mathcal{P}} \Sphi(\boldsymbol{\kappa})
\end{equation}
in which $J_1(\cdot)$ is a Bessel function of the first kind.
$\widetilde{\mathcal{P}}(\boldsymbol{\kappa})$ is the piston-removal
filter within square brackets in Eq. (\ref{eq:PRfilter}) for a circular pupil of diameter $D$. 
 
 Using the Parseval theorem, the residual (piston-removed) phase
 variance is defined by
  \begin{equation}\label{eq:reconstruction_error}
    \sigma^2_\text{Tot} \triangleq \int
     \widetilde{\mathcal{P}} \left\langle \left| \FTWF(\boldsymbol{\kappa}) - {\FTWF}^{\text{cor}}(\boldsymbol{\kappa})  \right|^2 \right\rangle
    \partial \boldsymbol{\kappa}
  \end{equation}
which is a function of $\{d,D,r_0,L_0,\sigma^2_\eta \}$,
the
actuator pitch, the telescope diameter, the atmosphere coherence
length, the outer scale and the measurement noise variance.

In the remainder we suppose that the DM corrects entirely for the
estimated phase, \textit{i.e.}
${\FTWF}^\text{cor}(\boldsymbol{\kappa}) =
\widehat{\FTWF}(\boldsymbol{\kappa})$ when the anti-folding filter is
applied \cite{correia14a} and implying
$\widetilde{\mathcal{F}}(\boldsymbol{\kappa}) = 1$, for $ |\boldsymbol{\kappa}| < 1/(2d)$ and
0 elsewhere.  The folding term in $\S$3.D of \textit{Correia et al (2014)}~\cite{correia14a} is not to be confused with
the aliasing term: it is related to the DM capability to correct for
frequencies above the cut-off frequency due to the cyclic pattern of the
DM actuators and shape of the influence functions \cite{ellerbroek05, correia14a}.

Equation (\ref{eq:reconstruction_error}) is expanded using
$\widehat{\FTWF}(\boldsymbol{\kappa}) = \FTR\FTS(\boldsymbol{\kappa})$
(the reconstructed phase)
and the general-purpose  measurement model in the Fourier domain
  \begin{equation}\label{eq:S_FT}
    \FTS\left(\boldsymbol{\kappa}\right) = \FTG
    \FTWF_{||}\left(\boldsymbol{\kappa}\right) + 
    \widetilde{\boldsymbol{\alpha}}\left(\boldsymbol{\kappa}\right) + 
    \widetilde{\boldsymbol{\eta}}\left(\boldsymbol{\kappa}\right),
\end{equation}
where $\FTG$ is a linear filter relating the wave-front $\FTWF$ to the slope
measurements $\FTS$ (which may include at a later stage detector integration and delay), $
\widetilde{\boldsymbol{\alpha}}\left(\boldsymbol{\kappa}\right)$ is
the aliasing term (equation 17 from \cite{correia14a}) acting as a
generalised measurement noise term and 
$\widetilde{\boldsymbol{\eta}}\left(\boldsymbol{\kappa}\right) $ is an
additive noise term \cite{correia14a}. 

Since the signal and noise processes are independent the cross terms
are zero and we can write
  \begin{equation}
\begin{split}
     \widetilde{\mathcal{P}}\average{\left| \FTWF(\boldsymbol{\kappa})
          - \widehat{\FTWF}(\boldsymbol{\kappa})  \right|^2}
   &  =
   \widetilde{\mathcal{P}}\average{\left|\tilde\varepsilon_\perp(\boldsymbol{\kappa})
     \right|^2} + \widetilde{\mathcal{P}}\average{\left|\tilde\varepsilon_{||}(\boldsymbol{\kappa})
     \right|^2}  \\ & 
= \average{|\FTWF_\perp|^2} \\ &\quad
    + \left|1 - \FTR \FTG\right|^2 \widetilde{\mathcal{P}} \average{\FTWF(\boldsymbol{\kappa})\FTWF(\boldsymbol{\kappa})^* }
    \\&\quad+ \mathbf{W}_\text{RA} \\ & \quad
    + \average{ \widetilde{\mathcal{P}} \left|\FTR \FTN\right|^2} 
\end{split}
  \end{equation}
  with $
  \average{|\FTWF_\perp|^2}$ the PSD of the fitting error (where we approximate
  $ \widetilde{\mathcal{P}}(\boldsymbol{\kappa}) = 1$ for $|\boldsymbol{\kappa}|>1/(2d)$).
  The term 
\begin{equation}
\left|1 - \FTR \FTG\right|^2 \widetilde{\mathcal{P}} \average{\FTWF(\boldsymbol{\kappa})\FTWF(\boldsymbol{\kappa})^* }=  \left|1 - \FTR \FTG\right|^2\Sphi'(\boldsymbol{\kappa})
\end{equation}
 is  the PSD of the static phase reconstruction error and 
  \begin{equation}
    \mathbf{W}_\text{RA}
= \widetilde{\mathcal{P}} \sum_{\mathbf{m}\neq 0}  \left|\FTR(\boldsymbol{\kappa}) \FTG(\boldsymbol{\kappa}+ \mathbf{m}/d)\right|^2 \Sphi(\boldsymbol{\kappa}  + \mathbf{m}/d)
  \end{equation}
  is the PSD of the static reconstructed aliasing error.  Finally  
\begin{equation}
\mathbf{W}_\eta = \average{ \widetilde{\mathcal{P}} \left|\FTR \FTN\right|^2}
\end{equation} 
is the PSD of the propagated noise.

\subsection{Dynamic case: closed-loop, $\tau\geq 0$, $\T_s > 0$}
\begin{figure}[htpb]
	\begin{center}
            \includegraphics[width=0.5\textwidth]{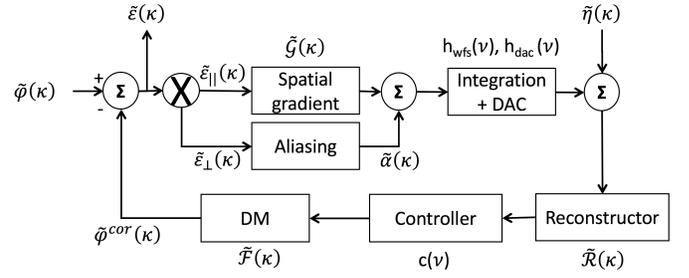}
	\end{center}
	\caption[]
	{\label{fig:loop}
     Block-diagram for closed-loop operation.  Variable definitions given in the
     text.}
\end{figure}

The dynamic case including loop filtering can now dealt with
straightforwardly. The block diagram is depicted in Fig. \ref{fig:loop}.
Equation \eqref{eq:errFunctionsExpanded} is still valid but now
$\varphivec^\text{cor}$ is a temporally-filtered version of
$\varphivec$.  
Here we introduce the Laplace
transform $\mathcal{L}\{\cdot\}$ which provides a
convenient treatment of the variables~\cite{oppenheim97}.

We will explicitly target closed-loop systems but, as will become apparent, 
the formulation accommodates the open-loop as a
special case. 

The closed-loop transfer function between the corrective and input
phase is given by \cite{madec99}
  \begin{equation}\label{eq:clTf}
    \mathbf{h}_\text{cor} (\kappavec, s) \triangleq \frac{\boldsymbol{\Phi}^\text{cor}}{\boldsymbol{\Phi}}(\kappavec, s) =
    \frac{\mathbf{h}_\text{ol}(\kappavec, s)}{1+\mathbf{h}_\text{ol} (\kappavec, s) }
  \end{equation}
where $s = 2i\pi\nu$ is the Laplace transform variable, with $\nu$ as the temporal
frequency in $seconds^{-1}$, such that $\boldsymbol{\Phi} (\boldsymbol\kappa, s)=
\mathcal{L}\{\varphivec(\boldsymbol\kappa , t)\}$, \textit{i.e.} each
spatial mode is temporally-filtered by functions
$\mathbf{h}_\text{cor} (\kappavec, s)$ where we explicitly show the
dependence on $\kappavec$ to include for instance a mode-by-mode loop gain. In Eq. \eqref{eq:clTf},
$\mathbf{h}_\text{ol} (\kappavec, s)$ is
the open-loop transfer function which includes pure loop time delays,
WFS integration time $T_s$ and other filtering aspects such as
the digital-to-analog conversion and the possible negative-feedback
regulator for the closed-loop case.  The mathematical form of these transfer
functions is given in detail in Sec.~\ref{sec:TF}.
At this point one can easily see that
to obtain the open-loop filtering it suffices to replace
$\mathbf{h}_\text{cor} (\kappavec, s)$ by $\mathbf{h}_\text{ol}
(\kappavec, s)$.


The transfer function of the noise relates the final phase error
to the initial noise, and describes the propagation of the noise through the loop
  \begin{equation}\label{eq:cl_ntf}
   \mathbf{h}_{\eta} (\kappavec, s) \triangleq
   \frac{\boldsymbol{\Phi}_\epsilon}{\boldsymbol{\Phi}_\eta}(\kappavec, s) =
   \frac{\mathbf{h'}_\text{ol} (\kappavec, s)}{1+\mathbf{h}_\text{ol} (\kappavec, s)} =  \frac{\mathbf{h}_\text{cor} (\kappavec, s)}{ \mathbf{h}_\text{wfs} (\kappavec, s)}
  \end{equation}
where $\mathbf{h'}_\text{ol} (\kappavec, s)$ 
is the noise transfer function, obtained from  $\mathbf{h}_\text{ol} (\kappavec, s)$ by
removing the WFS contribution (see Sec.~\ref{sec:TF}).

Therefore in closed-loop the term
\begin{equation}\label{eq:PRresisual}
 \widetilde{\mathcal{P}} \left\langle \left|
     \FTWF(\boldsymbol{\kappa}) -
     {\FTWF}^{\text{cor}}(\boldsymbol{\kappa})  \right|^2
 \right\rangle
\end{equation}
is computed using the temporally-filtered wave-fronts
\begin{equation}\label{eq:PhiCorrTemporallyFiltered}
{\FTWF}^{\text{cor}}(\boldsymbol{\kappa})  = \FTR \FTG \mathbf{H}_\text{cor}{\FTWF}(\boldsymbol{\kappa})
\end{equation}
The spatial transfer function at every spatial frequency is evaluated
from the temporal transfer function $\mathbf{h}_\text{cor}(\kappavec, s)_{|s=2i\pi\nu}$
\begin{equation}
\mathbf{H}_\text{cor}(\boldsymbol{\kappa}) = \mathbf{h}_\text{cor}\left(-2\pi i
\mathbf{v} \cdot \boldsymbol{\kappa}\right)
\end{equation}
where $\nu = \mathbf{v}_l \cdot\boldsymbol{\kappa}$ 
 and $\mathbf{v}_l = (v_x, v_y)_l$
 the wind velocity vector for the $l-th$ layer of the atmosphere. 
The minus sign within the function brackets is due to the
transformation from temporal to spatial variables using the frozen flow hypothesis, i.e. $\WF(\mathbf{x},t+\tau) =
 \WF(\mathbf{x} - \mathbf{v}\tau,t)$. Thus  positive time-delay $\tau$
is factored in as $\exp(-2\pi (-\mathbf{v} \tau) \cdot \boldsymbol{\kappa})$ which has a
positive exponent. 

For the general case with a stratified atmosphere, we now introduce
the functions
\begin{equation}\label{eq:H1}
\bar{\mathbf H}_{1} (\boldsymbol{\kappa}) = \sum_l r_{0,l}
\mathbf{h}_\text{cor}\left(-2\pi i \mathbf{v}_l \cdot
\boldsymbol{\kappa} \right) 
\end{equation}
and 
\begin{equation}
	\bar{\mathbf H}_{2} (\boldsymbol{\kappa}) = \sum_l r_{0,l} \left|\mathbf{h}_\text{cor}\left(-2\pi i \mathbf{v}_l
	\cdot \boldsymbol{\kappa} \right) \right|^2
\end{equation}
as weighted averages over the atmospheric layers,
$l=1\cdots L$,
each with it's own wind speed ($\mathbf{v}_l$) and fractional strength
($r_{0,l}$), insuring $\sum_l r_{0,l} = 1$. Plugging
Eq. \eqref{eq:PhiCorrTemporallyFiltered} into
Eq. \eqref{eq:PRresisual} and further developing the terms yields the
following relationships for the expected value of the filtered
wave-fronts  as
\begin{equation}
	\average{|\widehat{\FTWF}_{\parallel}|^2}
	\triangleq |\FTR \FTG|^2 \bar{\mathbf H}_2
	\average{|\FTWF_{\parallel}|^2} = |\FTR \FTG|^2 \bar{\mathbf H}_2 \Sphi 
\end{equation}
and    
\begin{equation}
	\average{\FTWF_{\parallel}
    	\widehat{\FTWF}_{\parallel}^*} \triangleq  \FTR \FTG \bar{\mathbf
    	H}_1 \average{|\FTWF_{\parallel}|^2} = \FTR \FTG \bar{\mathbf
    	H}_1 \Sphi
\end{equation}
the cross-term between the unfiltered
wave-front and its filtered version \cite{ellerbroek05, clare06}.

With these definitions we now have
\begin{align}
\average{|\FTWF_{\parallel} - \widehat{\FTWF}_{\parallel}|^2}  & =  \left(1 +
  |\FTR \FTG |^2 \bar{\mathbf H}_2 - \FTR \FTG \bar{\mathbf H}_1 -
                                               (\FTR \FTG
                                               \bar{\mathbf H}_1)^*\right)
                                               \Sphi
  \nonumber\\
& = \left(1 +
  |\FTR \FTG |^2 \bar{\mathbf H}_2 -
  2\Re\{\FTR \FTG \bar{\mathbf H}_1\}\right)
                                               \Sphi
\end{align}
The aliasing term is modified to account for the recursive filtering of
the loop 
  \begin{equation}
    \mathbf{W}_\text{RA}
= \widetilde{\mathcal{P}} \sum_{\mathbf{m}\neq 0}
\left|\FTR(\boldsymbol{\kappa}) \bar{\mathbf H}_1(\boldsymbol{\kappa})  \FTG(\boldsymbol{\kappa}+ \mathbf{m}/d)\right|^2 \Sphi(\boldsymbol{\kappa}  + \mathbf{m}/d)
  \end{equation}
as well as the noise propagation term:
\begin{equation}\label{eq:noisePropTerm}
 \mathbf{W}_\eta  = \average{ \widetilde{\mathcal{P}} \left|\FTR \FTN
     \mathbf{\bar{H}}_\eta (\boldsymbol{\kappa})\right|^2} = \widetilde{\mathcal{P}} \left|\FTR
     \right|^2 \sigma^2_\eta d^2 p
\end{equation}
where
\begin{equation}
\mathbf{\bar{H}}_\eta (\boldsymbol{\kappa}) =  \sum_l r_{0,l}\mathbf{h}_\eta (-2\pi i
\mathbf{v}_l \cdot \boldsymbol{\kappa})
\end{equation}
is the noise transfer computed from
 $\mathbf{h}_\eta (s)$ similarly to the calculation for the
 wave-front; on the right-hand side of Eq. \eqref{eq:noisePropTerm}
 $\sigma^2_\eta$ is the measurement noise variance, the
normalisation by $d^2$ comes from the 2D integration within the
correctable band and $p$is 
the noise propagation factor
\begin{equation}
p = 2T_s\int_0^{1/(2T_s)} |\mathbf{h}_\eta(s)|^2\dint s
\end{equation}
which provide amore accurate numerical evaluation accounting for the
loop rejection at very low temporal frequencies.

 It may be convenient to include some degree of uncertainty on our
 knowledge of both the 
 wind-speed and wind-direction \cite{ellerbroek05, clare06}  
taking an average over the range of possible values
\begin{equation}\label{eq:H1mod}
\bar{\mathbf H}_{1} (\boldsymbol{\kappa}) = \sum_l r_{0,l}
\int_{v_l-\Delta v_l}^{v_l+\Delta v_l}\int_{\theta_m}^{\theta_M}
\mathbf{h}_\text{cor}\left(-2\pi i \mathbf{v}_l \cdot
\boldsymbol{\kappa} cos(\theta)\right) \dint \theta \dint v.
\end{equation}
In Eq. \eqref{eq:H1mod} $v_l = |\mathbf{v}_l|$, $\Delta v_l$ is the wind-speed variation
during the integration and $\theta_M, \theta_m$ are the bounds for the
wind direction and
\begin{equation}
	\bar{\mathbf H}_{2} (\boldsymbol{\kappa}) = \sum_l r_{0,l} \int_{v_l-\Delta v_l}^{v_l+\Delta v_l}\int_0^{2\pi} \left|\mathbf{h}_\text{cor}\left(-2\pi i \mathbf{v}_l
	\cdot \boldsymbol{\kappa} cos(\theta)\right) \right|^2\dint \theta \dint v
\end{equation}
to give further flexibility evaluating the final error budgets.

We further note that computing $\left\langle \left| \FTWF(\boldsymbol{\kappa})
    - {\FTWF}^{\text{cor}}(\boldsymbol{\kappa})  \right|^2
\right\rangle$ can also be achieved by computing directly the error
functions $\left\langle \left| \tilde\varepsilon (\boldsymbol{\kappa})  \right|^2
\right\rangle$. In such case the residual error is computed as the
uncorrected wave-front filtered by the loop's rejection transfer
function. The closed-loop case is 
\begin{equation}\label{eq:cl_rtf}
    \mathbf{h}_\text{rej} \triangleq
    \frac{\boldsymbol{\Phi}_\epsilon}{\boldsymbol{\Phi}} =
    \frac{1}{1+\mathbf{h}_\text{ol}(s)} = 1- \mathbf{h}_\text{cor}(s)
  \end{equation}
leading as expected to a subtraction of a temporally-filtered version
of the wave-front from its original value as done repeatedly
throughout this presentation.

For off-axis observations, an anisoplanatic filter of the form
\begin{equation}
\widetilde{\mathcal{O}} = \sum_l r_{0,l} e^{2i\pi z_l\boldsymbol{\kappa}\cdot\boldsymbol{\xi}}
\end{equation}
is multiplied out by Eq. \eqref{eq:H1}, where $\boldsymbol{\xi}$ is the direction vector of the off-axis
star and $z_l$ is the altitude of the $l^{th}$ atmospheric layer. 

\subsection{Transfer-functions}
\label{sec:TF}
We assume the loop transfer function can be simplified to three components, namely the
WFSs, the digital-to-analog converters (DACs) and the controllers
\cite{roddier99}. This by no means precludes the future inclusion of
further
transfer functions representative of other devices in the AO loop.

The time-delays in the loop relate to the WFS integration and to the
RTC \textit{reconstruction and control} (R\&C) . The transfer functions for the
WFS and DAC are

\begin{equation}
\mathbf{h}_\text{wfs}(s) = \frac{1-e^{-2i\pi \nu T_s}}{s T_s}, \ \ \ \mathbf{h}_\text{dac}(s) = \frac{1-e^{-2i\pi \nu T_s}}{s T_s}
\end{equation}
where $T_{S}$ is the integration time. If we now put together the DAC
and WFS TFs, recalling the relationships  $\sinc(x) = \sin(\pi x)/(\pi x)$ and 
$(1-e^{-2i\pi \nu T_s})/(sT_s) = \sinc(T_s\nu) e^{-2i\pi\nu T_s/2}$. 
we get

\begin{equation}
\mathbf{h}_\text{wfs}(s) \times \mathbf{h}_\text{dac}(s) = \sinc^2(T_s\nu) e^{-2i\pi\nu T_s}
\end{equation}
which effectively introduces a one-step delay. The $\sinc^2$ term is
customarily dropped out since at low-frequencies (compared to $\frac{1}{T_{S}}$)
its magnitude is roughly equal to 1.

The open-loop transfer function  is consequently
\begin{equation}
\mathbf{h}_\text{ol}(s) = e^{-2i\pi qT_s\nu} \mathbf{c}(s)
\end{equation}
where  $\mathbf{c}(s)$ is the controller's transfer function for which
several options are discussed further below.

A discrete-time model is conveniently expressed in the Z-domain~\cite{oppenheim97}
\begin{equation}
\mathbf{h}_\text{ol}(z) = z^{-q}\mathbf{c}(z), \ \ \ z\triangleq e^{2\pi\nu T_s}
\end{equation}
 In a relatively standard case, the delays are taken to be
equivalent to 2 frames \cite{roddier99} for which we set $q=2$.

\section{Controller synthesis}\label{sec:controller-synthesis}

\subsection{State-space-based LQG controllers}

The derivation of the Linear-Quadratic-Gaussian (LQG) controller for AO applications has been presented in
detail elsewhere for both the infinitely fast DM response and
otherwise \cite{correia10a}. In this paper we take a shorter path and refer the reader to the bibliographic references on this particular subject for a more in-depth derivation. 

The discrete-time LQG regulator minimises the cost function
\begin{align}
J(\uvec) & = \lim_{M \rightarrow \infty}\frac{1}{M}\sum_{k=0}^{M-1} \left(\xvec^\T \mathbf{Q} \xvec  + \uvec^\T \mathbf{R} \uvec + 2 \xvec^\T \mathbf{S} \uvec \right)_k
\end{align}
where the weighting matrices $\{\mathbf{Q}, \mathbf{R},
\mathbf{S}\}$ are to be specified by developing an AO-specific
Strehl-optimal quadratic criterion 
subject to the state-space model
\begin{equation}\label{eq:state_space_model}
  \left\{\begin{array}{ccl}
     \xvec_{k+1}  &  = & \mathcal{A} \xvec_k  + \mathcal{B} \uvec_k  + \mathcal{V}  \boldsymbol{\nu}_k \\
     \svec_{k}& = & \mathcal{C}\, \xvec_k + \mathcal{D} \uvec_{k-d} + \etavec_k
    \end{array}\right.
\end{equation}
where $\xvec_k$ is the state vector that contains the wave-front
coefficients (or any linearly related variables), $\svec_{k}$ are the
noisy measurements provided by the WFS in the guide-star (GS) direction;
$\boldsymbol{\nu}_k$ and $\etavec_k$ are spectrally white,
Gaussian-distributed state excitation and measurement noise
respectively. In the following we assume that
$\CovMat_{\boldsymbol{\nu}} = \average{\boldsymbol{\nu}
  \boldsymbol{\nu}^\T}$,  $\CovMat_{\mathbf{\boldsymbol{\eta}}} =
\average{\boldsymbol{\eta} \boldsymbol{\eta}^\T}$ are known and that
$\CovMat_{\mathbf{\boldsymbol{\eta,\nu}}} = \average{\boldsymbol{\eta}
  \boldsymbol{\nu}^\T} = 0$. The definition of the model parameters
\{$\mathcal{A},\mathcal{B}, \mathcal{C}, \mathcal{D}$\} is obtained
from physical modelling of the system dynamics and response functions \cite{correia10a}.

We also take the case where $\mathcal{D}$ is a delay operator that
acts upon $\uvec_k$. For instance if $\mathcal{D} = z^{-2}$, then with
abuse of mathematical notation $\mathcal{D}\uvec_k \rightarrow
\mathcal{D}(\uvec_k) = \uvec_{k-2}$. 

We next provide a model for a distributed Kalman-filter (DKF) that
assumes a basis of complex exponential functions
\cite{massioni11, poyneer05},  creating oscillatory modes
\begin{align}
f_{\kappavec} (\rhovec) = \exp\left(-\frac{2i\pi \rhovec\cdot\kappavec}{N}\right)
\end{align}
that when discretised are used to construct a 
 mode-by-mode complex-valued
state-space model as shown below. Note that this model is computed on
mode-per-mode fashion, leading to many parellel calculations of
small-scale, independent controllers (hence the designation \textit{distributed}) 

We define the state as a concatenation of complex-valued variables
\begin{equation}
{\xvec}_{k} (\kappavec) = 
\left(\begin{array}{c}
 \widetilde\varphivec^1_k\\ \cdots\\ \widetilde \varphivec^L_k\\
        \widetilde\phivec_{k}\\\widetilde\phivec_{k-1} 
\end{array}\right)(\kappavec)
\end{equation}

\begin{equation}
{\mathcal{A}}({\kappavec}) = 
\left(\begin{array}{ccc}
\left[\widetilde{\mathbf{P}}(\kappavec) \right]& \0 & \0  \\
\text{diag}\left(\left[\widetilde{\mathbf{P}}(\kappavec)\right]\right) & \0 & \0 \\
\0 & 1 & 0 \\
\end{array}\right)
\end{equation}
with $\left[\widetilde{\mathbf{P}}({\kappavec})\right] $ a
diagonal matrix concatenating the $L$-layers translation coefficients
such that 
\begin{align}
\alpha_l(\kappavec) & = \xi \exp(-2 i\pi\nu_l)
\end{align}
where $\xi $ is a real-valued scalar smaller than unity that ensures overall
controller stability \cite{correia10a} and 
with the temporal frequency vector 
 $\nu_l = -\mathbf{v}_l \cdot\boldsymbol{\kappa}$.  The operation
 $\text{diag}\left(\left[\widetilde{\mathbf{P}}({\kappavec})\right]
 \right)= [\alpha_1,\cdots,\alpha_L]$ extracts and concatenates in a
 row-vector the diagonal coefficients of
 $\left[\widetilde{\mathbf{P}}({\kappavec})\right]$, representing the
 accumulation of wave-fronts at the telescope pupil-plane,
 i.e. $\widetilde\phivec_{k}=\sum_l \widetilde \varphivec^l_k$.

Matrices 
\begin{equation}
{\mathcal{B}}({\kappavec}) = 
\left(\begin{array}{c}
\0 \\ 0\\0 
\end{array}\right)
\end{equation}
\begin{equation}
{\mathcal{V}}({\kappavec}) = 
\left(\begin{array}{c}
\I_{L\times L} \\ \0\\\0
\end{array}\right)
\end{equation}
\begin{equation}\label{eq:C}
{\mathcal{C}}({\kappavec}) = 
\left(\begin{array}{ccc}
\0 & 0  & \FTR \FTG ({\kappavec}) 
\end{array}\right) 
\end{equation}
\begin{equation}\label{eq:D}
{\mathcal{D}}({\kappavec}) = 
\left(\begin{array}{c}
-\FTR \FTG ({\kappavec})
\end{array}\right)
\end{equation}

are chosen following the disturbance-rejection mode of the AO
dynamical control problem \cite{correia10a}.

The choice of model in Eqs. \eqref{eq:C} and \eqref{eq:D} where the
Kalman filter takes into account the spatial filtering provided by
$\FTR \FTG$ gives us the possibility to chose spatial reconstructors
$\FTR$ to mitigate aliasing \cite{correia14a}. 

The treatment of the aliasing term provided here is formulated in its spatial
dimension instead of temporal \cite{poyneer10}. The filter that is
obtained is therefore more compact and lends itself more suitably to
real-time implementation (in the discrete version). 
This choice is
different from the one in \textit{Poyneer et al} \cite{poyneer10}
where an augmented state is suggested to accommodate the aliasing
modes; this would have the downside of increasing the state dimensions
considerably, by a factor of 5 if only the direct neighbours whose
frequencies fold-in to the AO correctable band are
considered (i.e. $\kappavec + \mathbf{m}/d$ with $\mathbf{m} =
\{0,1\}$, $\mathbf{m} = \{0,-1\}$, $\mathbf{m} = \{1,0\}$, $\mathbf{m}
= \{-1,0\}$) and by a factor of 9 if the cross terms where to be
considered.

The implementation of the KF involves a real-time state update and 
prediction equations 
\begin{subequations}\label{eq:single_rate_StatSA_RToperations}
\begin{align}
  \widehat{{\xvec}}_{k|k}(\kappavec) & =
                                       \widehat{{\xvec}}_{k|k-1}(\kappavec)
  \nonumber \\& + \widetilde{\mathcal{H}}_\infty\left(\FTS_k(\kappavec) -
      \mathcal{C} \widehat{{\xvec}}_{k|k-1}(\kappavec) + \mathcal{D} \uvec_{k-q} \right) 
\\
  \widehat{ {\xvec}}_{k+1|k}(\kappavec) & =
                                       {\mathcal{A}}\widehat{
                                                   {\xvec} }_{k|k}
                                       (\kappavec) 
\end{align}
\end{subequations}
where $\mathcal{H}_\infty$ is the asymptotic Kalman gain computed from
the solution of an estimation Riccati equation \cite{correia10a}. 

In order to evaluate the AO residuals with the DKF, its transfer
function must be developed. This is accomplished next.

\begin{property} \textbf{LQG transfer functions.}
  
  Let the general controller $\mathbf{c}(\z) $ be a transfer function
  such that
  \begin{equation}
            \mathbf{u}(\z) = \mathbf{c}(\z) \mathbf{s}(\z).
          \end{equation}
where $\mathbf{u}(\z)$ is the output (commands) and $\mathbf{s}(\z)$
is the inputs (measurements).

          For an LQG using the estimator version, $\uvec_k =
          -\K \widehat{\xvec}_{k|k}$
          the LQG controller transfer function 
          is
          \begin{equation}
\begin{split}
    \mathbf{c}(\z)  & = - \left\{\I + \K \Lambda_{e} \right. \\ &
    \quad \quad \left. \left[z^{-1}
        (\I-\Hkal\Cd) \Bd + \z^{-d}\Hkal\mathcal{D}\right] \right\}^{-1} \K \Lambda_{e} \Hkal,
\end{split}
  \end{equation}
  where 
  \begin{equation}
    \Lambda_{e} = \left(\I - \z^{-1}\Ad + \z^{-1} \Hkal \Cd \Ad \right)^{-1}.
  \end{equation}
Demonstration is provided in Appendix \ref{sec:lqgTFs}.  \propend
\end{property}

\begin{figure}[htpb]
	\begin{center}
            \includegraphics[width=0.5\textwidth]{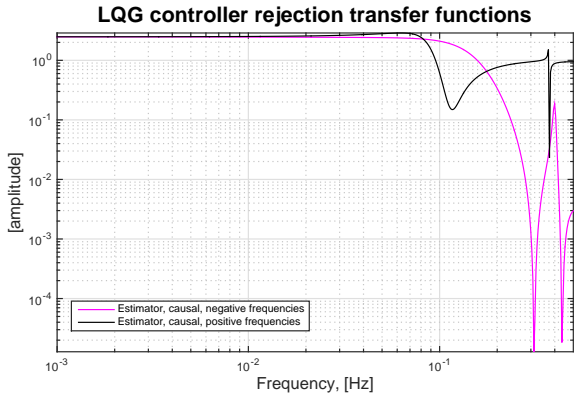}
	\end{center}
	\caption[]
	{\label{fig:distLqgTfs_3LayerAtm}
       Distributed Kalman filter transfer functions (positive and
       negative temporal frequencies shown) for spatial mode
       $\kappa_x=\kappa_y=2$ and a 3-layer
       atmosphere.}
\end{figure}

Figure \ref{fig:distLqgTfs_3LayerAtm} illustrates the rejection
transfer function for a DKF controller encompassing a 3-layer atmosphere
with $v_x=[5; 10; 7]\,m/s$, $v_y= [0; 0; 0]\,m/s$,
       $T_s=1\,s$, leading to 3 notches at
       temporal frequencies $\nu = [-0.3125 ;  -0.6250 ;  -0.4375]\,Hz$. 

\subsection{Integral action controllers}
The commonly used integrator can be formulated from the state-space
model \cite{kulcsar09} but we take a shorter path stating only the
more convenient
final formulation as follows 
\begin{equation}
\mathbf{c}(s) = \frac{g}{1-e^{-2\pi \nu T_s}}
  \end{equation}
with $g$ the integrator gain chosen to ensure the best performance
within the
stability bounds.

        \section{Shack-Hartmann-based \textit{classical} and \textit{extreme} AO }\label{sec:SHmodelling}

A comprehensive presentation of the treatment that follows can be found in full in
\cite{correia14a}.   Here we jump straight to the main result which is
the filter formulation of the Shack-Hartmann gradients~\cite{roddier99,
  rigaut98, hardy98} in the spatial-frequency domain.

Let the SH-WFS measurements $\Sl(\mathbf{x},t)$ be given by the
  geometrical-optics linear model  
  \begin{equation}\label{eq:s_Gphi}
    \Sl(\mathbf{x},t) = \G\WF(\mathbf{x},t) + \Noise(\mathbf{x},t),
  \end{equation}
  relating aperture-plane wave-fronts $\WF(\mathbf{x},t)$ to measurements through
  $\G$, a phase-to-slopes linear operator over a bi-dimensional space indexed by $\mathbf{x} = (x,y)$ at time $t$;  $\Noise(\mathbf{x},t)$ represents white
  noise due to photon statistics,
  detector read noise and background photons. Both $\WF$ and $\Noise$ are zero-mean functions of Gaussian probability distributions and known covariance matrices: $\Sigma_\phi$ and
  $\Sigma_\eta$ respectively. Additive, Gaussian-distributed noise is assumed both temporally and
  spatially uncorrelated. Variances are evaluated following formulae in \cite{thomas06}
for shot photon and read noise on centre-of-gravity centroiding of SH spots
\begin{equation}
\sigma^2_{\Delta\phi,ph}=\frac{\pi^2}{2 ln(2)}\frac{1}{n_{ph}}\left(\frac{N_T}{N_D}\right)^2 
\end{equation}

\begin{equation}
\sigma^2_{\Delta\phi,det}=\frac{\pi^2}{3}\frac{\sigma^2_{ron}}{n_{ph}^2}\left(\frac{N_S^2}{N_D}\right)^2 
\end{equation}
where
\begin{itemize}
\item $n_{ph}$ is the number of photons per sub-aperture and exposure
\item $\sigma_{ron}$ is the standard deviation of the read-out noise per
  pixel and per exposure
\item $N_T$ is the spot's FWHM in number of pixels
\item  $N_D$ is the sub-aperture's FWHM in number of pixels
\item  $N_S$ is the linear size of the window where the CoG is computed
\end{itemize}

The total number of photons is computed from $n_{ph}=N_{ph}\times T_s
\times throughput \times d^2$, where $N_{ph}$ is the number of
photons/$meter^2$/second emanating from the source with a zero-point of $N_{ph}=8.5\times10^9$
in the R band ($\lambda_\text{WFS}=0.64\mu m$). Furthermore
we have taken $N_S = N_D = N_T = 2$. 

Developing  the linear operator $\G$, it can be shown that
  \begin{equation}
    \G = \mathsf{III}\left(\frac{\mathbf{x}}{d}\right) \times \left[\Pi\left(\frac{\mathbf{x}-1/2}{d}\right) \otimes\Pi\left(\frac{\mathbf{x}}{\mathbf{v}T_s}\right) \otimes \nabla\right],
  \end{equation}
  where $\otimes$ is a 2-dimensional convolution product, $\times$ is a point-wise multiplication and $\Pi (\cdot)$ 
  is the unit rectangle separable function 
  \begin{equation}
    \Pi(\mathbf{x})   \triangleq \left\{\begin{array}{ll}
        1&  \text{if}\,\, |x|\leq 1/2 \land |y| \leq 1/2 \\
        0& \text{otherwise}
      \end{array}
    \right. .
  \end{equation}
$\mathsf{III}$ is a comb function that represents the sampling
process, \textit{i.e.} picking sample measurements at the corners of
each sub-aperture following the Fried measurement geometry.

 Using common transform pairs for the individual operations (see e.g.~\cite{oppenheim97}), the Fourier
  representation of the measurement
  yields
  \begin{equation}\label{eq:FT_G}
    \FTG = \TFd\{\G\} = \mathsf{III}\left(\boldsymbol{\kappa} d\right) \otimes [\FTP\left(\boldsymbol{\kappa} d\right) \times e^{i \pi d \boldsymbol{\kappa}} \times \widetilde{\nabla} \times \FTP\left(\boldsymbol{\kappa} \mathbf{v}T_s\right)],
  \end{equation}
  where $\TFd\{\cdot\}$ represents the continuous Fourier transform,  $\boldsymbol{\kappa} = (\kappa_x,\kappa_y) \in \mathbb{R}^2$ 
  is the frequency vector and
  $i\triangleq\sqrt{-1}$.  
  In Eq. \eqref{eq:FT_G} the gradient-taking
 part is  
  \begin{equation}
    \widetilde{\nabla}(\boldsymbol{\kappa}) = 2 i \pi d \boldsymbol{\kappa} = 2 i \pi d [\kappa_x, \kappa_y],
  \end{equation}
and the spatial averaging process is
  \begin{align}
    \FTP\left(\boldsymbol{\kappa} d\right)& = \sinc (d\boldsymbol{\kappa}) 
  \end{align}
 The temporal integration function reads
  \begin{equation}
    \FTP\left(\boldsymbol{\kappa} \mathbf{v}T_s\right) = \sinc\left(\boldsymbol{\kappa} \mathbf{v}T_s\right)
  \end{equation}
on account of the frozen-flow assumption made earlier.
We note that $\FTG$ only acts as a
  filtering function for functions within the AO correctable band~\cite{correia14a}. 

The Fourier-domain representation of Eq. (\ref{eq:s_Gphi}) is then
  \begin{equation}\label{eq:S_FT}
    \FTS\left(\boldsymbol{\kappa}\right) = \FTG
    \FTWF_{||}\left(\boldsymbol{\kappa}\right) +
    \widetilde{\boldsymbol{\alpha}}\left(\boldsymbol{\kappa}\right) + \widetilde{\boldsymbol{\eta}}\left(\boldsymbol{\kappa}\right),
  \end{equation}
where $\{\cdot\}_{||}$ represents variables within a selection of
frequencies $|\kappavec_{x,y}|\leq 1/(2d)$, thus obtaining
 \begin{equation}
  \FTG \triangleq 2\, i\, \pi\, d\, \boldsymbol{\kappa}\, \FTP\left(\boldsymbol{\kappa}d \right)\, e^{i \pi d \boldsymbol{\kappa}}\, \FTP\left(\boldsymbol{\kappa} \mathbf{v}T_s\right)\, \FTWF_\parallel\left(\boldsymbol{\kappa} \right) 
\end{equation}
and an aliasing term seen as the folded measurements beyond the AO
cut-off frequency over spatial modes within the band
\begin{equation}
\widetilde{\boldsymbol{\alpha}}\left(\boldsymbol{\kappa}\right) = \sum_{\mathbf{m}\neq 0}  \FTG \left(\boldsymbol{\kappa} + \mathbf{m}/d \right) \FTWF\left(\boldsymbol{\kappa} + \mathbf{m}/d \right)
\end{equation}
\section{Limiting post-coronagraphic contrast under predictive
  closed-loop control}\label{sec:postCoronaConstrast}

With the previous developments, we can now investigate the impact of
optimal time and spatial filtering when compared to standard least-squares,
integrator-based controllers. 

Using the Taylor expansion of the point-spread function  (PSF) in \cite{perrin03} but including
phase-only effects, for the sake of the discussion we assume that a coronagraph is capable of completely
removing the two first terms (the diffraction and pinned-speckles
term), leaving a halo term that is essentially the power spectrum of
the wave-front aberrations \cite{macintosh05}.  

Under these assumptions, the post-coronagraphic intensity relates to variance by
\begin{equation}\label{eq:ContrastFromVariance}
i_{\text{\tiny PC},\kappa}(\rhovec) = \sigma^2_\kappa \left[ i(\rhovec - \kappa
\lambda) + i(\rhovec + \kappa
\lambda)\right]
\end{equation}
where $ \sigma^2_\kappa$ is the variance of spatial mode indexed by
frequency $\kappa$ and $i_\text{\tiny PC}(\rhovec)$ is the post-coronagrahic intensity pattern
(image) in the focal plane.

From Eq. \eqref{eq:ContrastFromVariance} it can
readily be seen that the post-coronagraphic image is proportional to the
post-AO residual PSD. A more complete treatment can be found in
\cite{herscovici17}, which provides an analytic expression for
long-exposure post-coronagraphic images through turbulence but was not
adopted here. 

Post-coronagraphic long-exposure contrast is defined as
\begin{equation}\label{eq:contrast}
c(\rhovec) = \frac{i_\text{\tiny PC}(\rhovec)}{i(0)}
\end{equation}
where $i(\rhovec)$ is the AO-corrected
PSF with $i(\rhovec=0)=S_\text{\tiny LE}$ is
scaled by the long-exposure Strehl-ratio $S_\text{\tiny LE}$. It is computed as the
ratio of the AO-corrected optical transfer function
(OTF) integrated over all frequencies by the
diffraction-limited OTF. PSFs are computed from the
power-spectral density (PSD), using well-established
relationships. First we compute covariance functions from the
PSDs, using the Wiener–-Khintchine theorem. Computation of the spatial
structure function ensues, from which the OTF is determined. The PSF is found from the Fourier transform of
the OTF.
\begin{figure}[htpb]
	\begin{center}
            \includegraphics[width=0.25\textwidth]{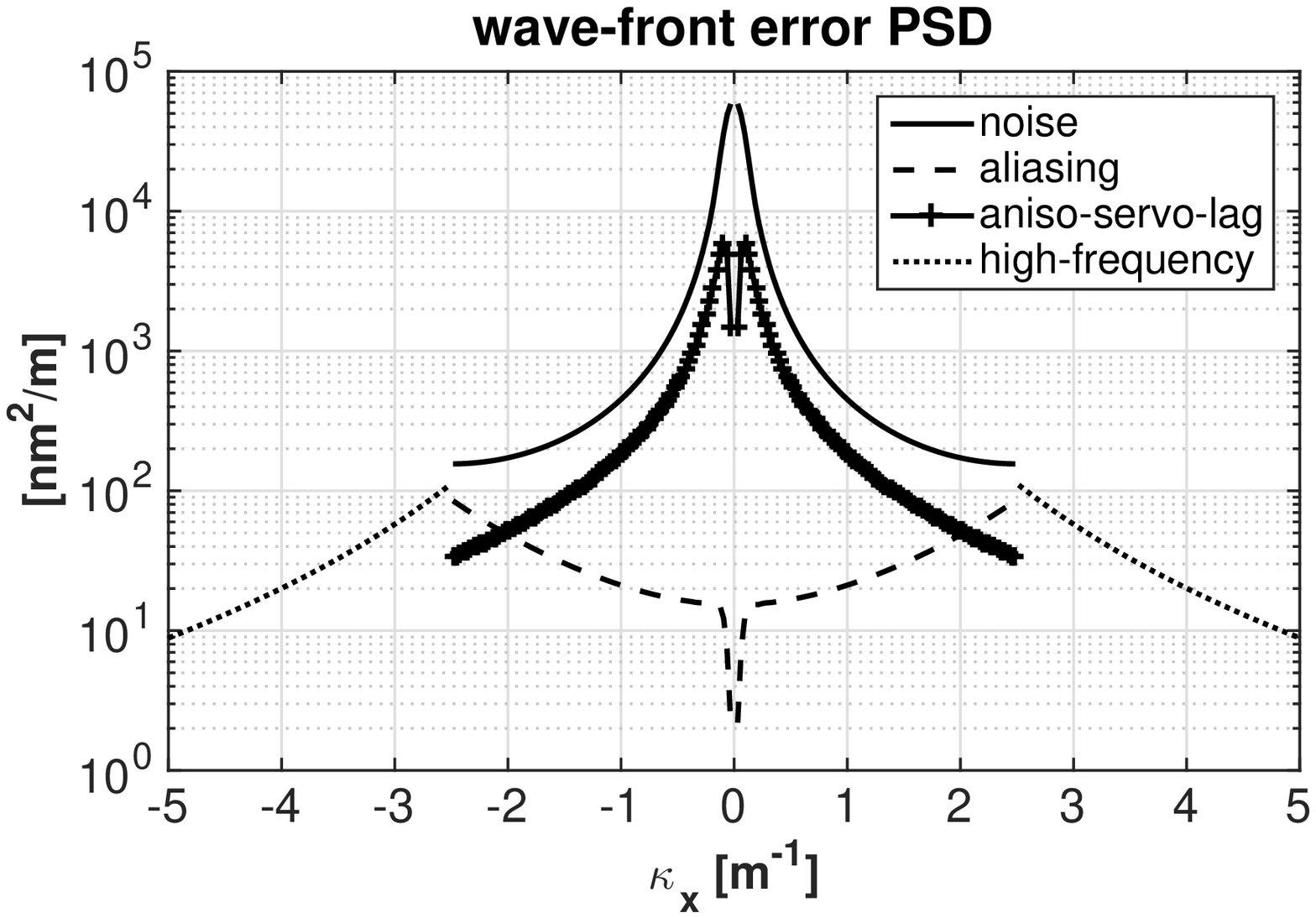}\includegraphics[width=0.25\textwidth]{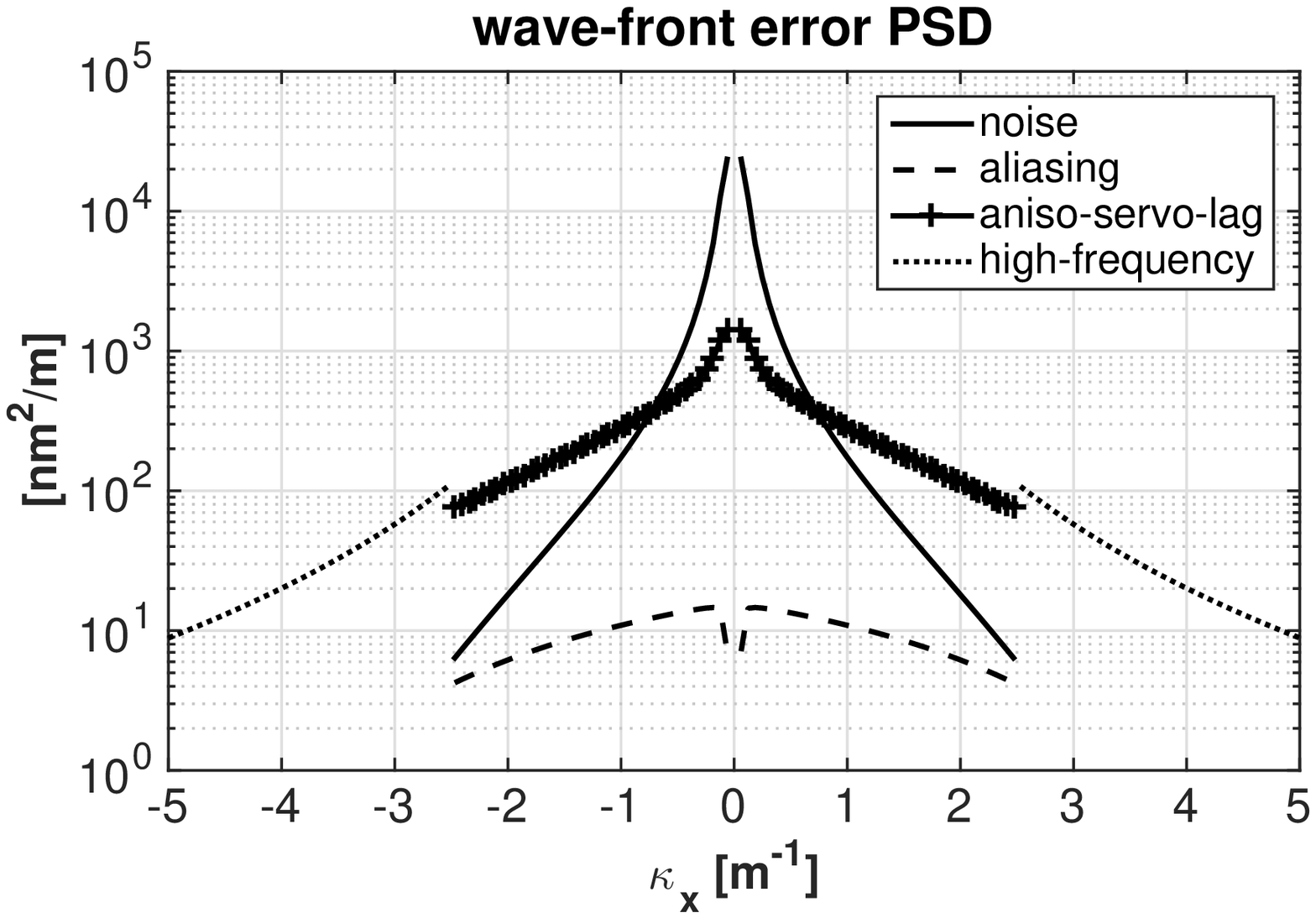}
	\end{center}
	\caption[]
	{\label{fig:slabs_vlt9layersAtmLQGvsInt}
    	PSD slabs for comparison of error profiles.  Calculated for a multilayer atmosphere
    	and a SPHERE-like system (see text for parameters). Left: integrator controller and LS filter; bottom: DKF
	controller with MV filter. Note the rebalancing of the error
        terms, with the DKF essentially trading noise propagation for
        aniso-servo-lag error, with a dramatic decrease in aliasing.}
\end{figure} 

Figure \ref{fig:slabs_vlt9layersAtmLQGvsInt} shows radial slabs of
 the residual spatial PSDs for the 4 individual error terms considered. One can
 see the enhanced rejection of errors close to the AO cut-off
 frequency for both
 aliasing and noise, leading to overall better performance in terms of
 wave-front residual and contrast across the AO-corrected field. The case considered is a 8\,m
telescope with 40 sub-apertures SH-WFS observing a magnitude 12 star in the R-band through
a stratified atmosphere with 9 layers with $r_0=15.5\,cm$ and outer
scale $L_0=25\,m$ with AO cut-off $\pm1/(2d) =\pm 2.5m^{-1}$. 

In passing, we note that when properly scaling the
residual PSDs according to \eqref{eq:ContrastFromVariance} gives the
limiting post-coronagraphic for the ideal coronagraph. We add to the
work of \cite{guyon05} where the term $"C_2"$ accounts for the effect
of WFS noise and time lag, the filtering of the closed-loop controller
and the propagation of aliasing for the case of a stratified atmosphere.


\begin{figure}[htpb]
	\begin{center}
            \includegraphics[width=0.25\textwidth]{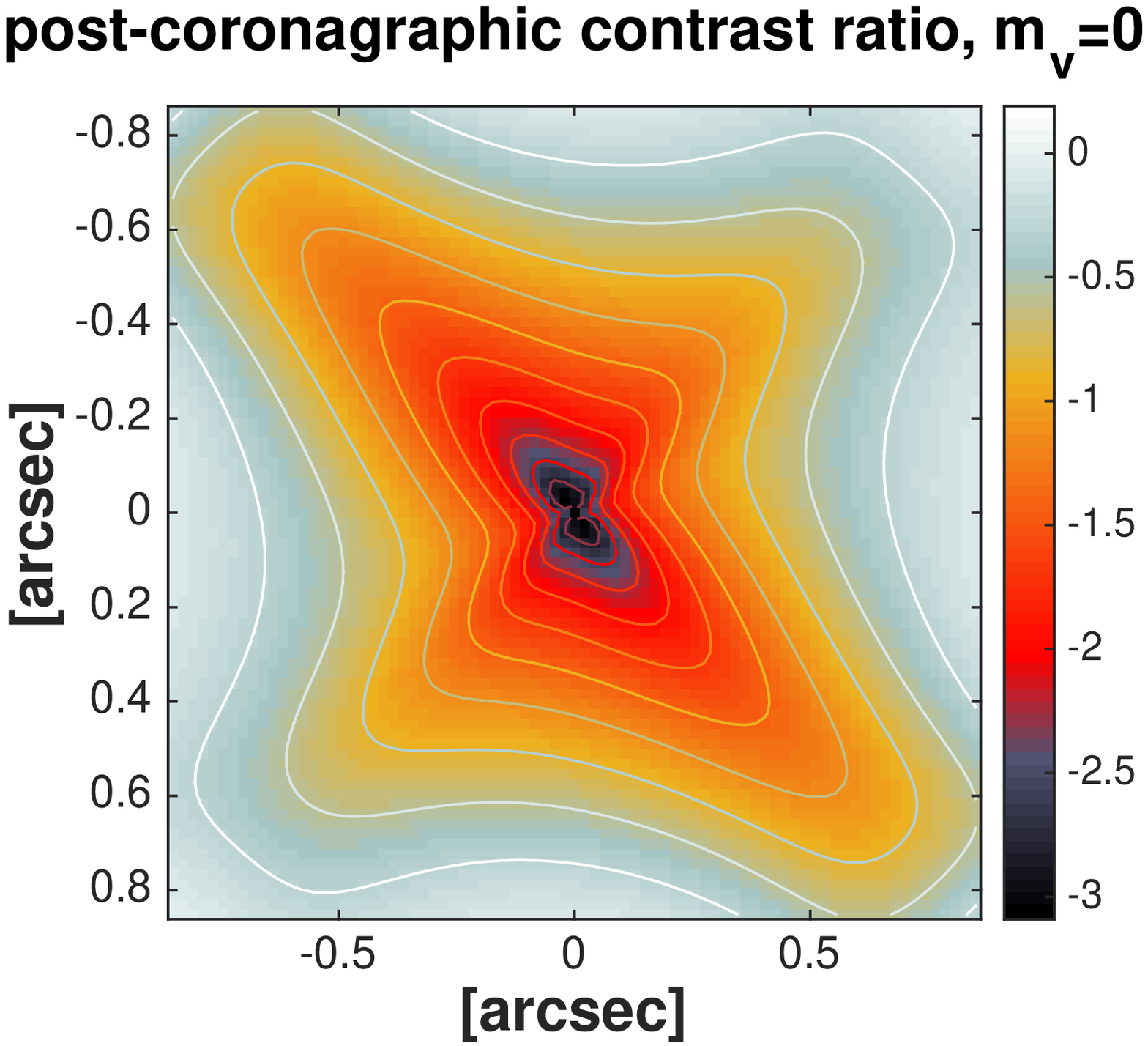}\includegraphics[width=0.25\textwidth]{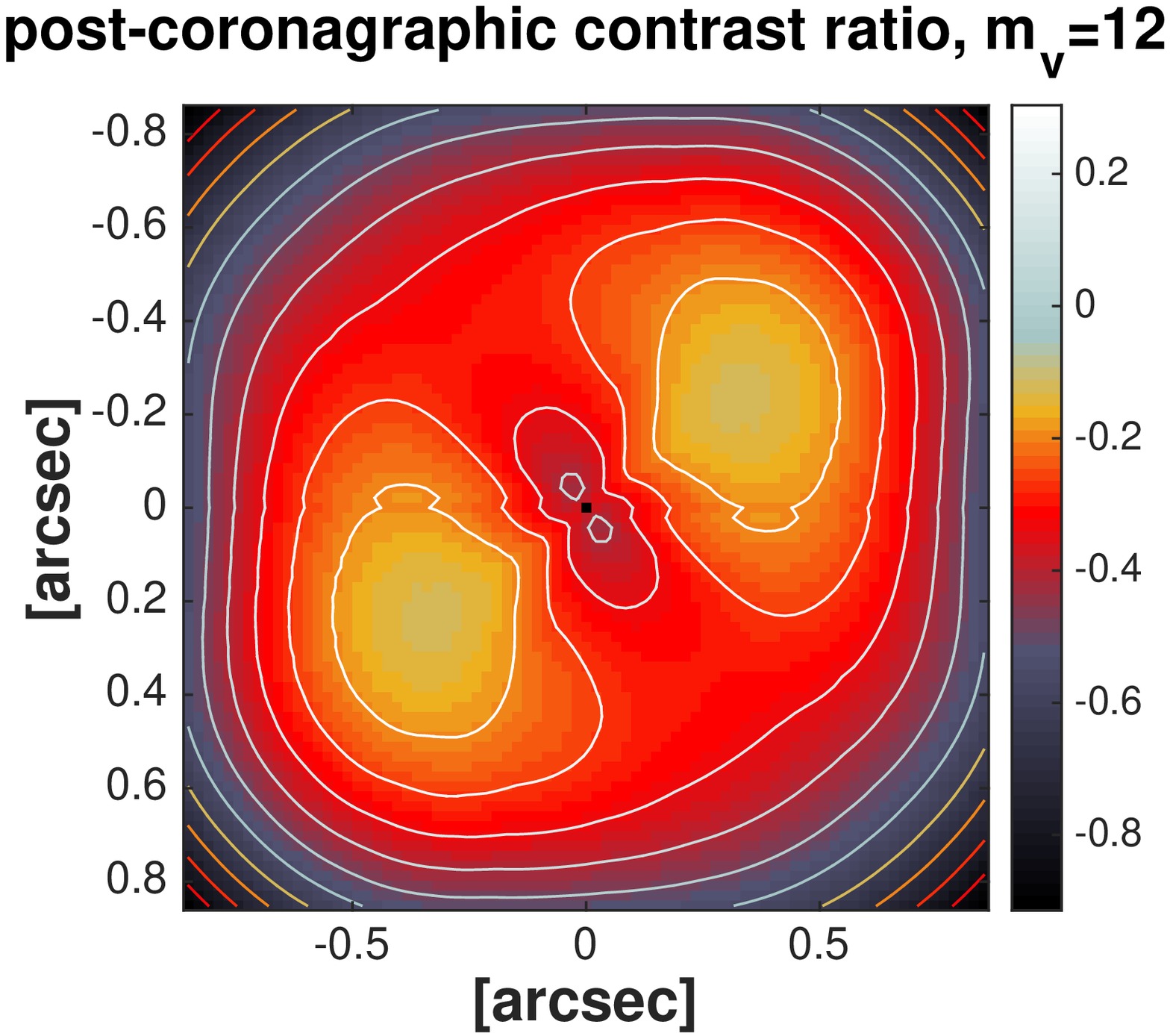}
	\end{center}
	\caption[]
	{\label{fig:sphereContrastEnhancement}
   Plots (log scale) illustrating the improvement in contrast using a DKF controller with a MV reconstructor, versus
   an integrator controller with a LS reconstructor. (Left:) magnitude
   0 star, (Right:) magnitude 12 star.
    }
\end{figure}

Figure \ref{fig:sphereContrastEnhancement} shows ratios of constrast
obtained from Eq. \eqref{eq:contrast} for the integrator controller with
least-squares phase reconstruction against the DKF with
minimum-variance reconstructor, \textit{i.e.} $c_\text{DKF}(\rhovec)/c_\text{INT}(\rhovec)$. On the bright
   star end improvement of 2 orders of magnitude could be attained at
   2-5 $\lambda/D$ for existing high-contrast imagers on a 10\,m-class telescope whereas one order of
   magnitude can still be achieved at separations closer to the AO
   control radius on a 12th magnitude star, on accound of better
   aliasing handling by the DKF. 

\section{Keck's NIRC2 performance and post-coronagraphic contrast enhancement}\label{sec:examples}

In this section we have focused on the W. M. Keck--II NIRC2 AO system
as a baseline configuration. Table \ref{tab:KeckBaseline} gathers the main
parameters used to produce these results.

\begin{figure}[htpb]
	\begin{center}
            \includegraphics[width=0.5\textwidth]{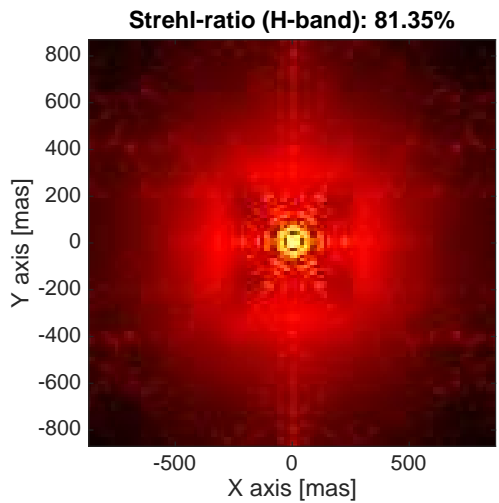}
	\end{center}
	\caption[]
	{\label{fig:psfTotal_KeckMultiLayerLS}
    Face-on PSF for a DKF controller with the Anti-Aliasing reconstruction filter .
    }
\end{figure}

In Table~\ref{tab:KeckResults} the error terms for different control
systems are summarised, 3 instances of an integrator controller
and a Discrete Kalman Filter (DKF).
\begin{table}[t]
\caption
{Keck--II simulation parameters.}
\vskip 2mm
\begin{center}
\begin{tabular}{ll}
\hline \hline
{\bf Telescope} & \\
D & 11.25\,m \\
throughput & 50\%\\
\hline
{\bf Guide-star} & \\
zenith angle & 0 deg \\
magnitude & 8\\
\hline
{\bf Atmosphere} & \\
$r_0$ &  16\,cm \\
$L_0$ & 75 m \\
Fractional $r_0$ & [0.517, 0.119, 0.063, \\ & 0.061, 0.105, 0.081, 0.054]\\
Altitudes & [0,0.5, 1, 2, 4, 8, 16]\,km\\
wind speeds & [6.7, 13.9, 20.8, 29.0, \\ & 29.0, 29.0, 29.0]\,m/s \\
wind direction & [$0,\pi/3,-\pi/3, -\pi,$ \\ & $-4/3\pi, -\pi/6, \pi/8$]\,rad \\
\hline
{\bf Wavefront Sensor} & \\
Order & 20$\times$20 \\
RON & 4.5\,e$^-$ \\
$N_{pix}$ & 12 \\
$f_{sample}$ & 500\,Hz \\
$\lambda_\text{WFS}$ & 0.64\,$\mu$m \\
Centroiding algorithm & thresholded CoG\\
\hline
{\bf DM} & \\
Order & 21$\times$21 \\
\hline
{\bf AO loop} & \\
pure delay  & $\tau_\mathsf{lag}=$4\,ms \\
\hline
{\bf Imaging Wavelength} & \\
$\lambda_\text{IM}$ (H-band) & 1.65\,$\mu$m \\
\hline
\end{tabular}
\end{center}
\label{tab:KeckBaseline}
\end{table}
\begin{table*}[htpb]
\caption[] {\label{tab:keckResults}
    Error budget for Keck--II model in nm\,rms using the parameters in
    Table \ref{tab:KeckBaseline}. The 4 errors terms considered are
    outlined in \S \ref{sec:WF_alias_noise_error}. }
  \begin{center}
    \begin{tabular}{c c c c c c c}
      \hline \hline
      Error term  	& Int+LS 	& Int+MV  & Int+AA 	& DKF+LS 	& DKF+Rigaut	&DKF+AA\\ \hline
      Aliasing      	& 55.4  	& 55.1 	& 43.7	        & 56.0 		& 55.7 		& 44.1\\
      Aniso-servo-lag  	& 61.0  	& 60.9	& 66.2 	        & 1.1 		& 1.5			& 32.0\\
      WFS noise          & 13.6  	& 13.6	& 12.9     	& 18.9  	& 18.9		& 18.1\\
      Total                  & 83.5      & 83.3      & 80.4       & 59.1 &  58.8  &  57.4\\
      Total + Fitting   & 141.2  	&141.0 	&139.4  	& 128.3       	& 128.2 		& 127.5       \\
      SR @1.65$\mu m$ & 77.2\% & 77.2\% & 77.9\%& 81.1\%& 81.1 \% &    81.4\%\\
     \hline \hline
    \end{tabular}
  \end{center}
  \label{tab:KeckResults}
\end{table*} 
The use of the DKF results in a significant drop in the aniso-servo-lag error,
as expected, becoming negligible compared to the other errors present
in the system (under the assumption of perfect knowledge of wind
parameters). Figure \ref{fig:psfTotal_KeckMultiLayerLS} shows the PSF obtained with
a DKF with anti-aliasing reconstructor.

\begin{figure*}[htpb]
	\begin{center}
          {   
          \begin{tabular}{ccc}
            \includegraphics[width=0.33\textwidth]{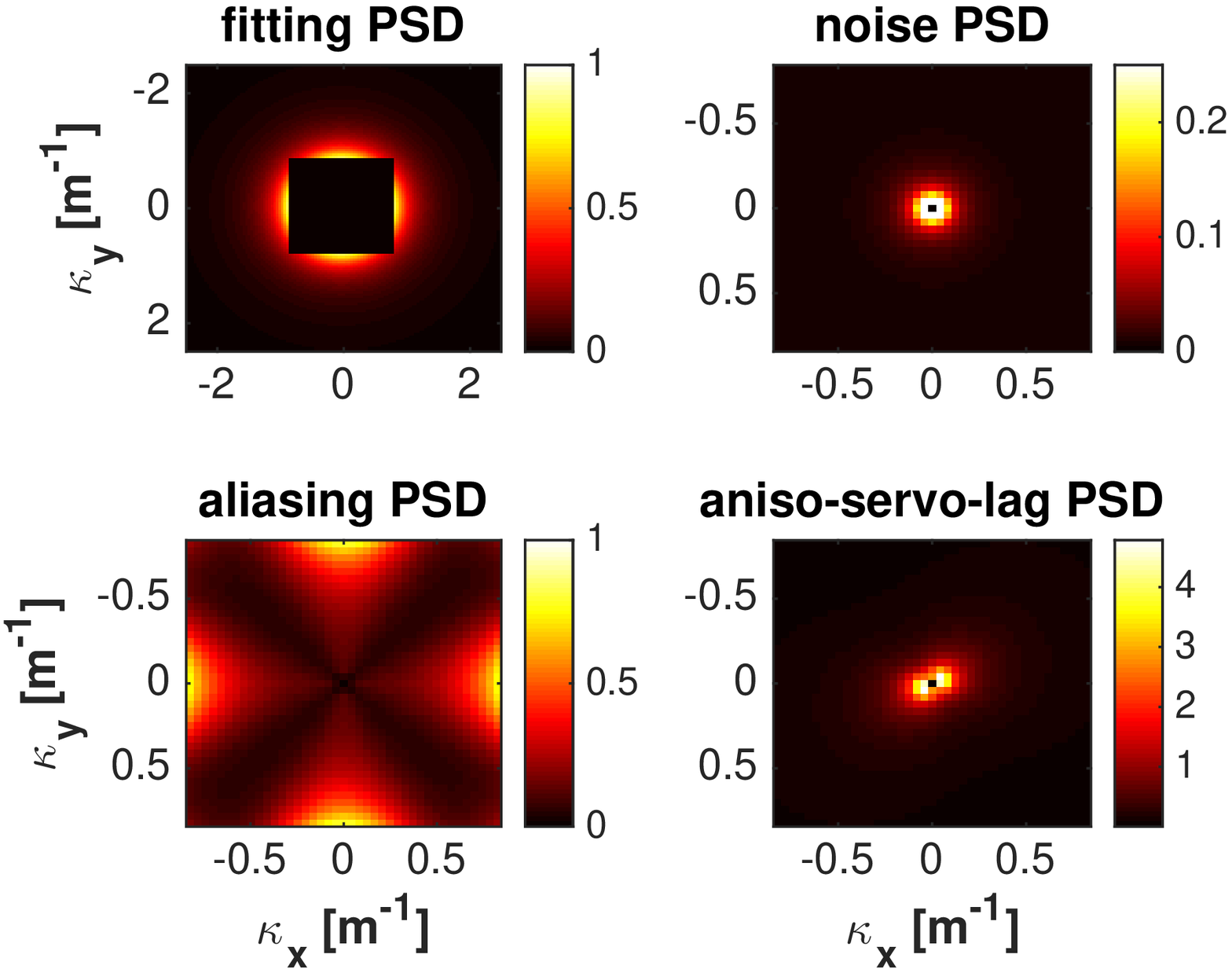}& 
              \includegraphics[width=0.33\textwidth]{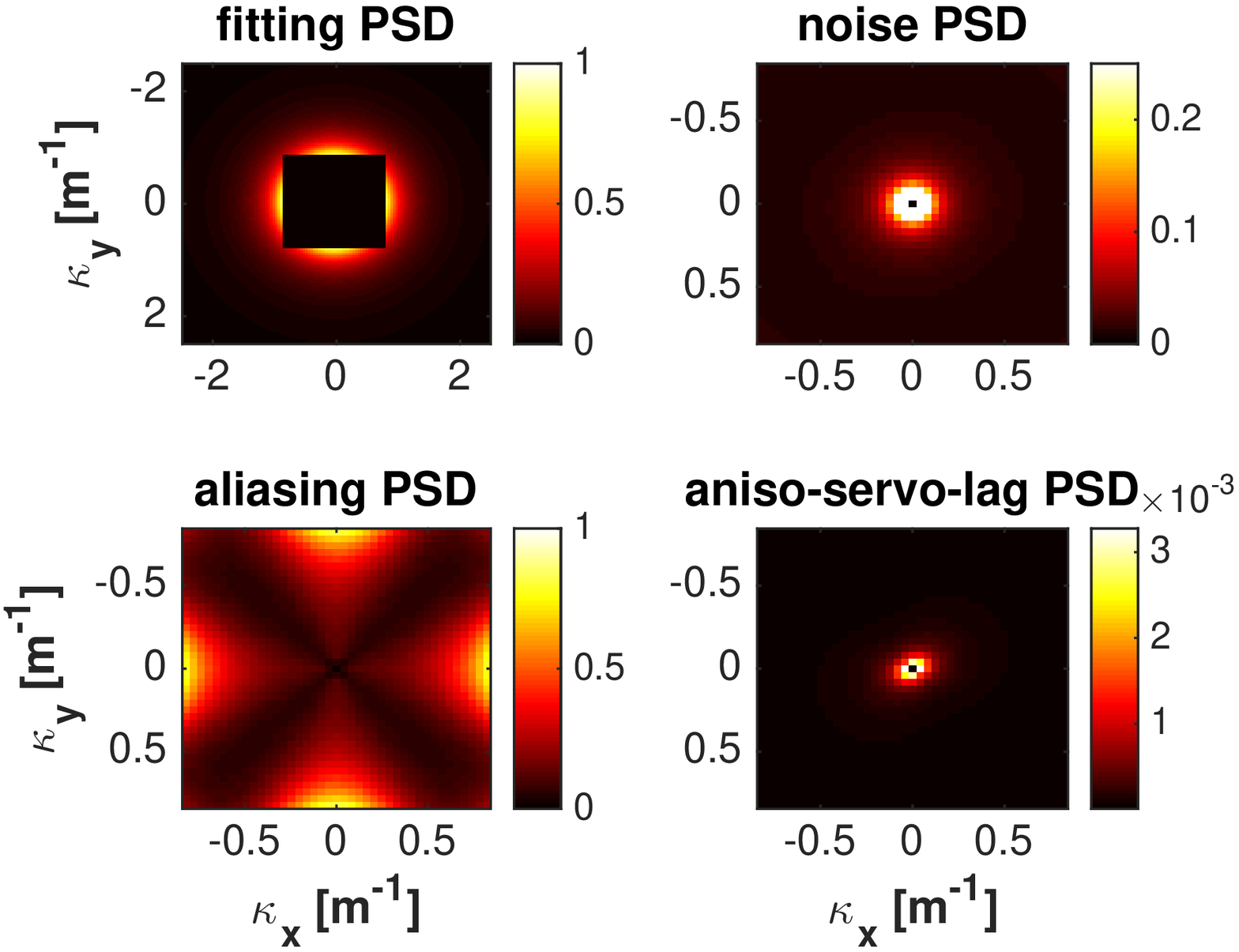}& 
               \includegraphics[width=0.33\textwidth]{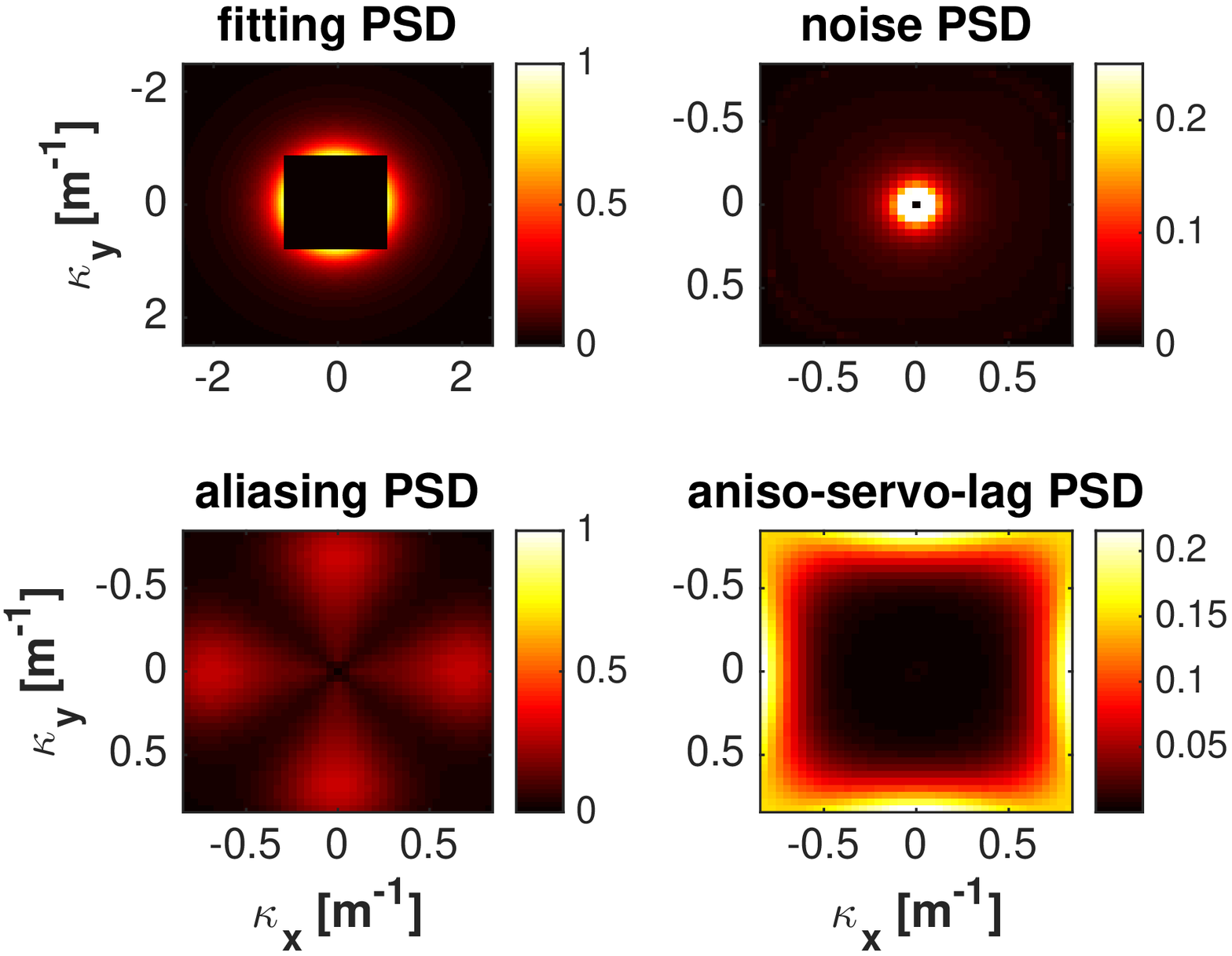}
           \end{tabular}
         \label{fig:psdi_KeckMultiLayerInt}
         } \\
         {
         \begin{tabular}{ccc}
           \includegraphics[width=0.33\textwidth]{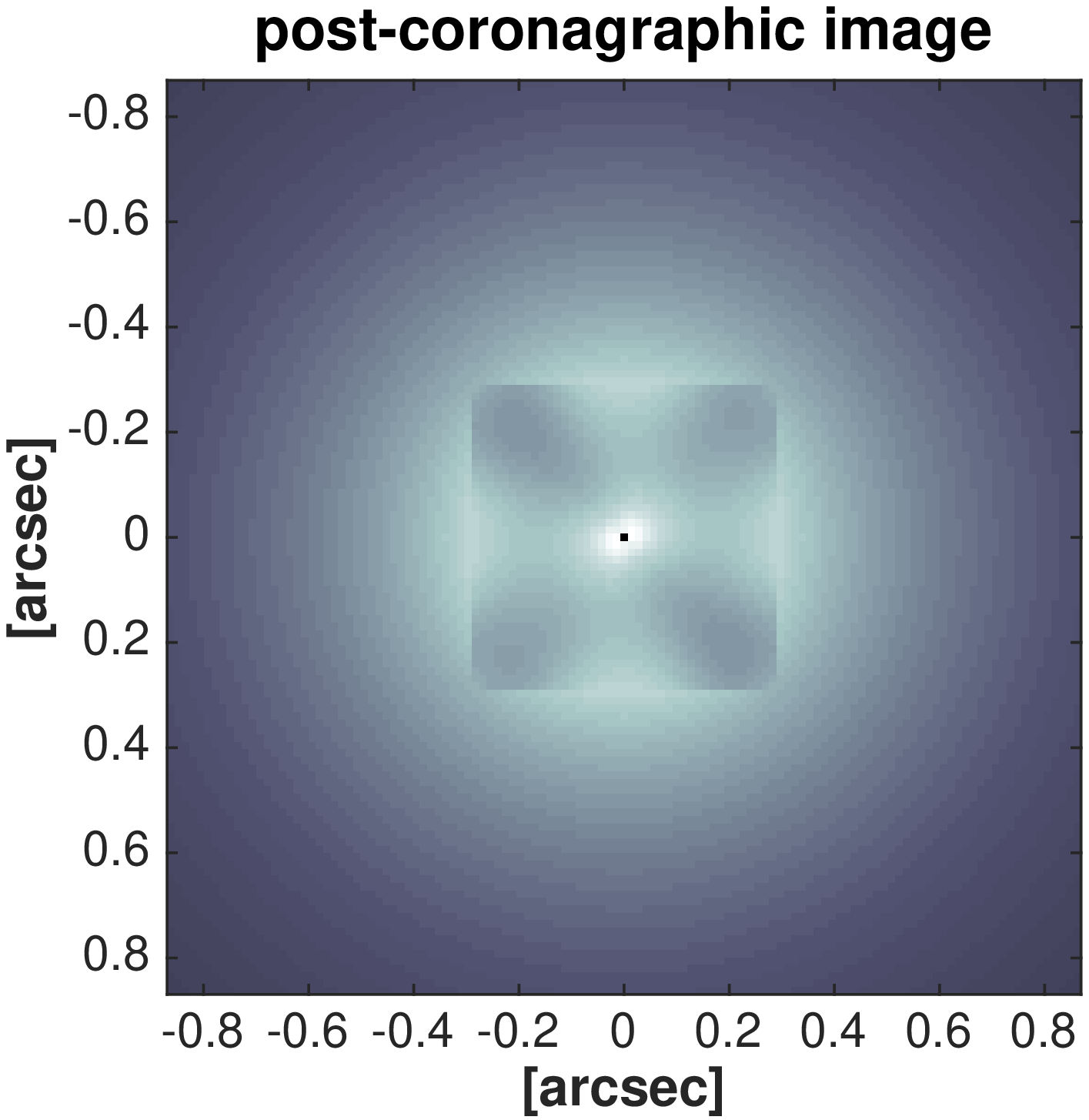}&
            \includegraphics[width=0.33\textwidth]{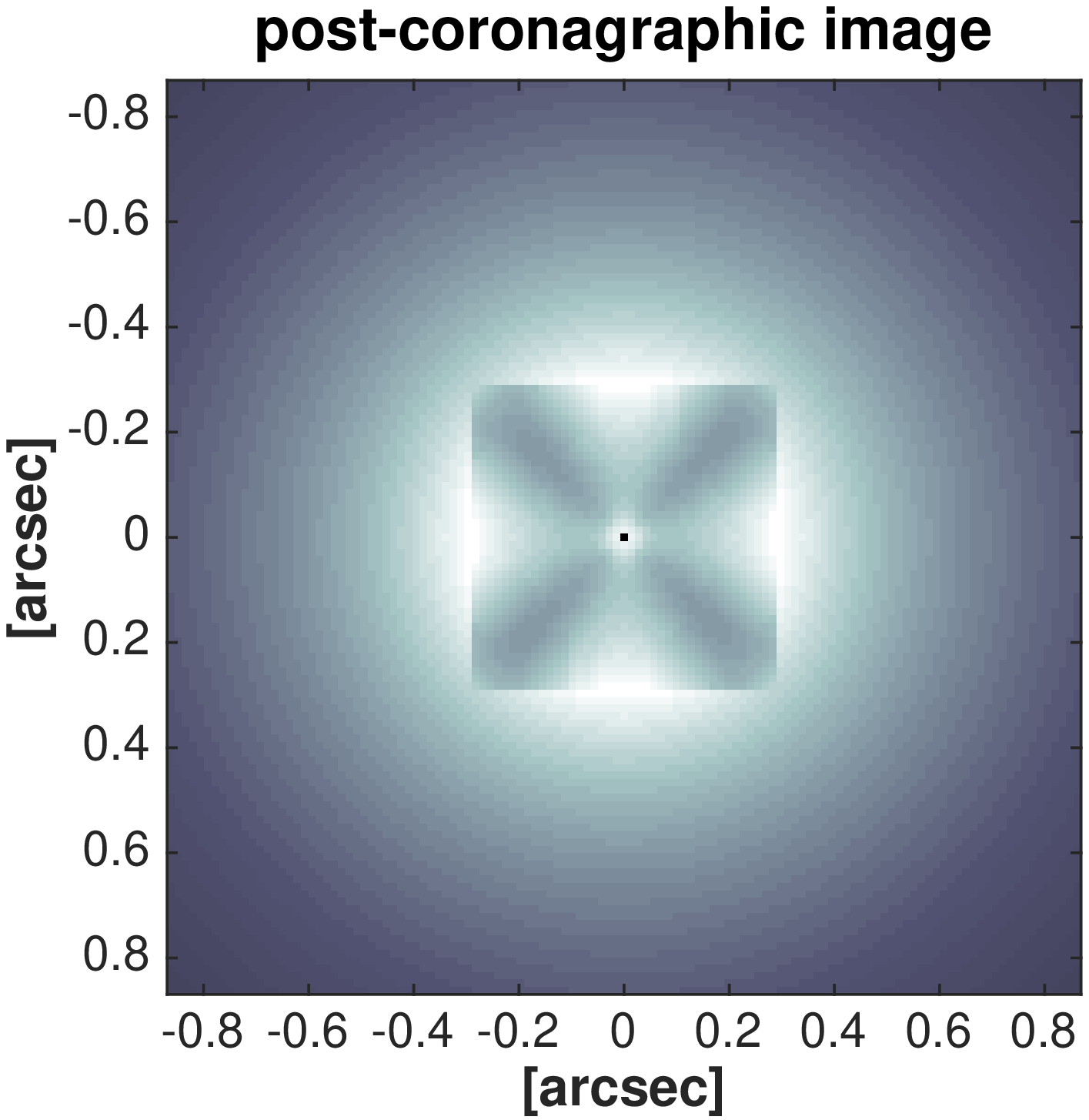}&
            \includegraphics[width=0.33\textwidth]{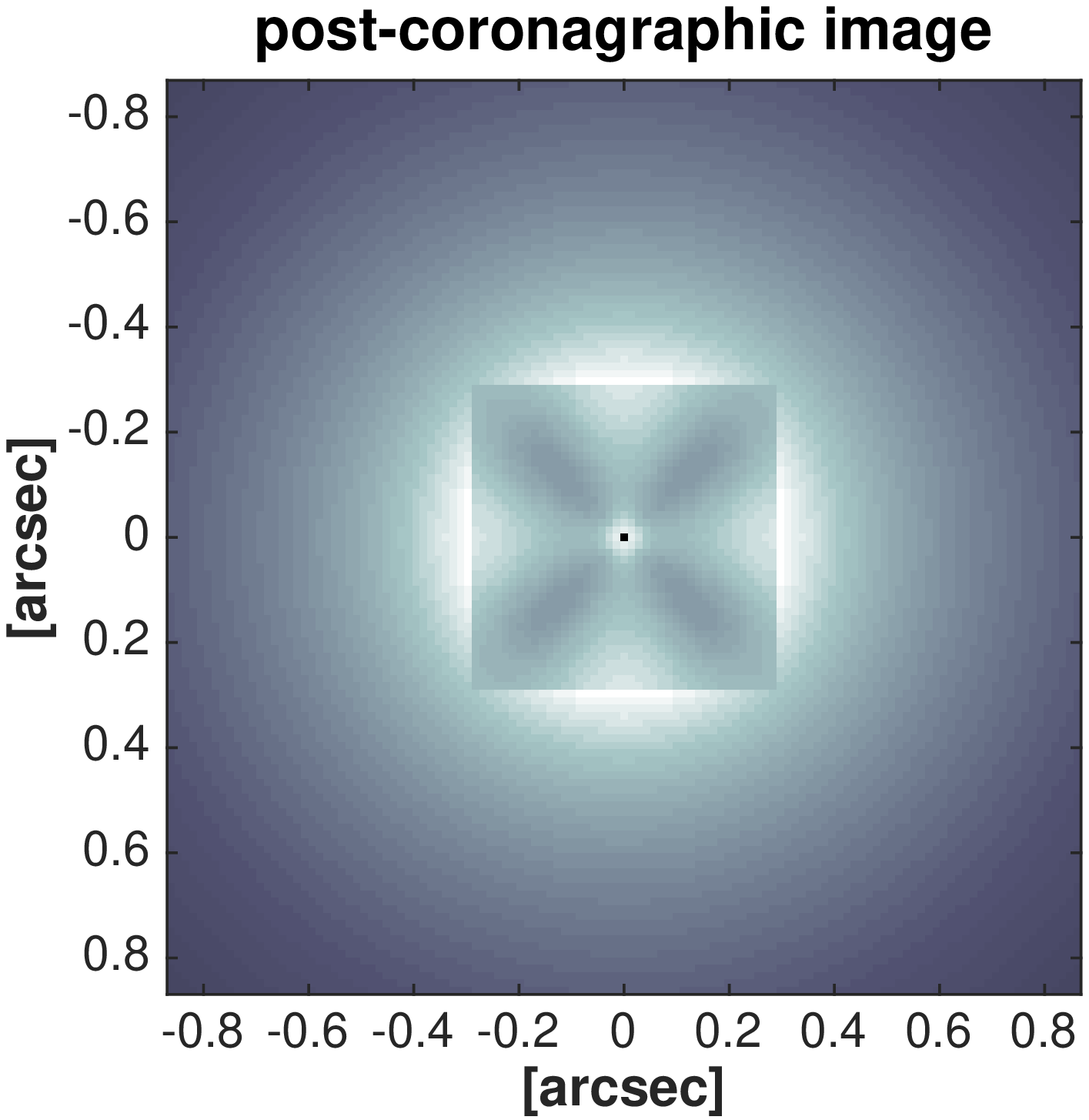}
         \end{tabular}         
         } 
	\end{center}
	\caption[]
	{\label{fig:psdi_KeckMultiLayerLQG}
     Residual PSDs for a multi-layer atmosphere with Keck--II parameters using different controllers. 
     (Top:) the face-on PSDs of the fitting (NW), noise (NE), 
     aliasing (SW) and aniso-servo-lag (SE); (bottom); the ideal post-coronagraphic 
     image of a point-source on account of the AO phase-only errors. (Left:) Integrator with a least-squares
     filter. (Centre:) DKF controller coupled to a least-squares
     filter. (Right:) DKF controller coupled to an anti-aliasing filter.}
\end{figure*}

Figure~\ref{fig:psdi_KeckMultiLayerLQG} depicts the 2D
PSDs for each error term, used here as a 1st-order approximation of an ideal post
coronagraph image as per Eq. \eqref{eq:ContrastFromVariance} for the different controllers considered~\cite{sauvage16}.
The fitting PSD depicts the high frequency
error, or \emph{fitting error}, and in all cases is zero within the correctable band and
then follows the trend of the turbulent phase PSD, $\propto \boldsymbol{\kappa}^{-11/3}$.
All other errors affect only the correctable band 
($|{\boldsymbol{\kappa}}|\leq \frac{1}{2d}$) and only this region is depicted.
The noise PSD follows a trend $\propto \boldsymbol{\kappa}^{-2}$
resulting from the spatial integration through
$\widetilde{\mathcal{R}}(\boldsymbol{\kappa})$. 

The aliasing PSD in Fig.~\ref{fig:psdi_KeckMultiLayerLQG} demonstrates the classic shape of
an aliased post-coronagraphic image, with the largest errors towards
the edges of the correctable band in a cross-like pattern.
This can be somewhat mitigated by using an anti-aliasing filter (Fig.~\ref{fig:psdi_KeckMultiLayerLQG}--right).
Finally the servo-lag PSD shows two distinct lobes in the integrator case
(Fig.~\ref{fig:psdi_KeckMultiLayerLQG}-left), the direction
of which is correlated with the predominant wind speed and direction
weighted by the
relative layer
strength (typically the elongation will follow the wind at the ground), with the overall scale related to
the wind speed compared with the AO system's frame-rate.  High wind-speeds
or large time delays will incur larger errors here.  
The final images show the sum of these error as an estimation of the
post-coronagraphic PSF. 

Comparing the DKF to the integrator, one can
readily observe the tremendous reduction in servo-lag error. With the
DKF with perfect knowledge of the system and atmospheric
parameters (see entry \textit{aniso-servo-lag} in Table \ref{tab:keckResults})
it is practically reduced to null. The estimated PSF images
show the AO-corrected PSF over a field 5 times larger than
the AO-corrected field. For this example the strong winds at the lower
layers
induce a stretching of the PSF with an integrator controller
(as seen in Fig.~\ref{fig:psdi_KeckMultiLayerLQG}).
The ideal case when the
turbulence layer strengths, speeds and directions are known is shown
as a limiting case. The image within the correction-band becomes much
sharper when the DFK is utilised. 

The improvement in contrast is illustrated in Fig.~\ref{fig:contrast}.
This demonstrates the significant improvement of the DKF over an
integrator controller, nearly a factor of 10 improvement in contrast close to the
halo (see magnitude and noise levels on Table \ref{tab:KeckResults}).
When the DKF is coupled with the anti-aliasing reconstructor, it can
still improve contrast in some areas, particularly so where aliasing
is stronger at the edges of the AO control region at the expense of
lower constrast in other regions.

The anti-aliasing filter used in this example does not include the effect of
temporal filtering on the aliasing errors (only the open loop aliasing propagation)
and hence does not improve performance over the entire correction
band. Optimisation of the aliasing regularisation term would require
iterative non-linear minimisation routines to chose the best term on
both temporal and spatial frequencies; we leave these developments for
a subsequent analysis.


\begin{figure*}[htpb]
	\begin{center}
          \begin{tabular}{cc}
            \includegraphics[width=0.4\textwidth]{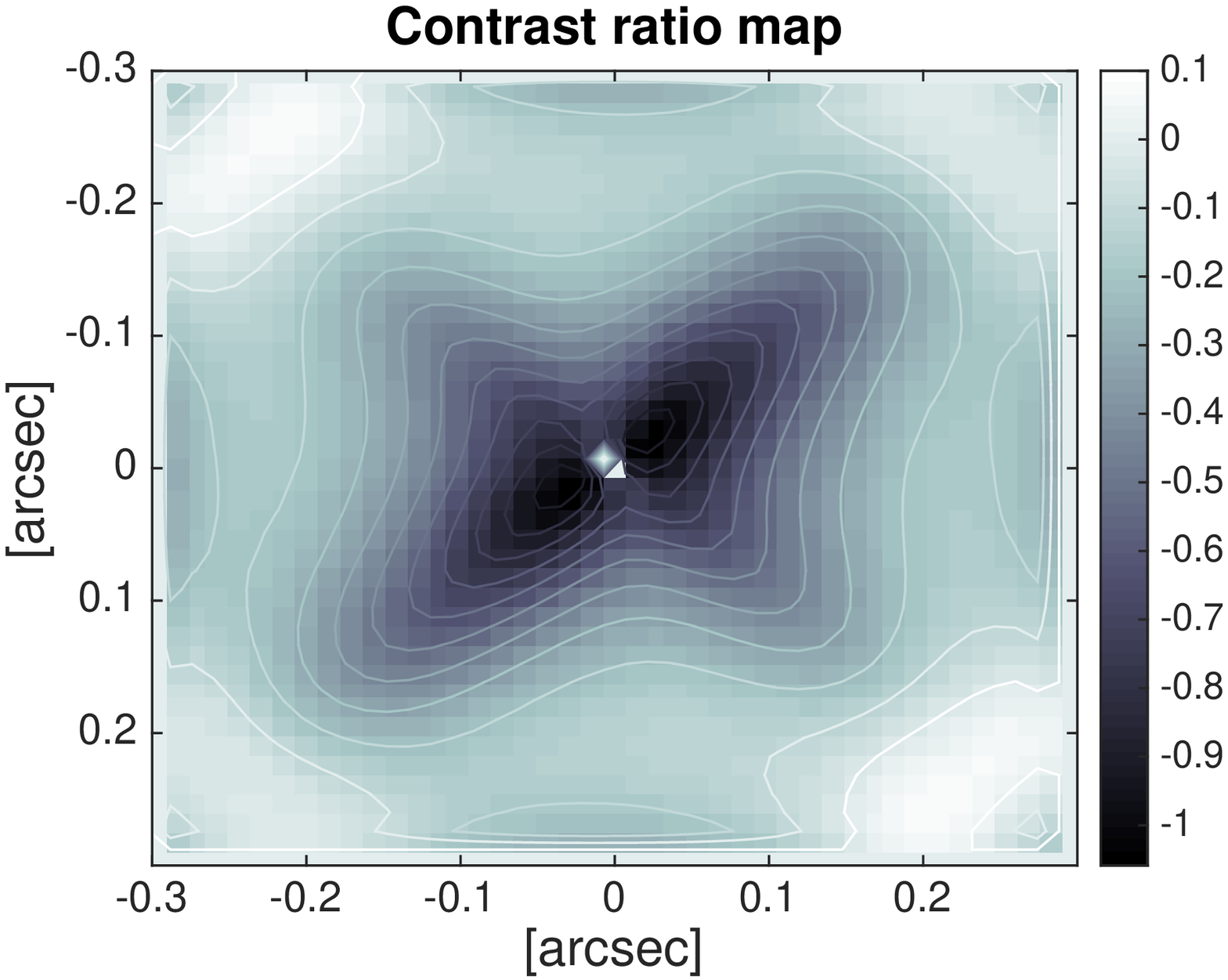}&
            \includegraphics[width=0.4\textwidth]{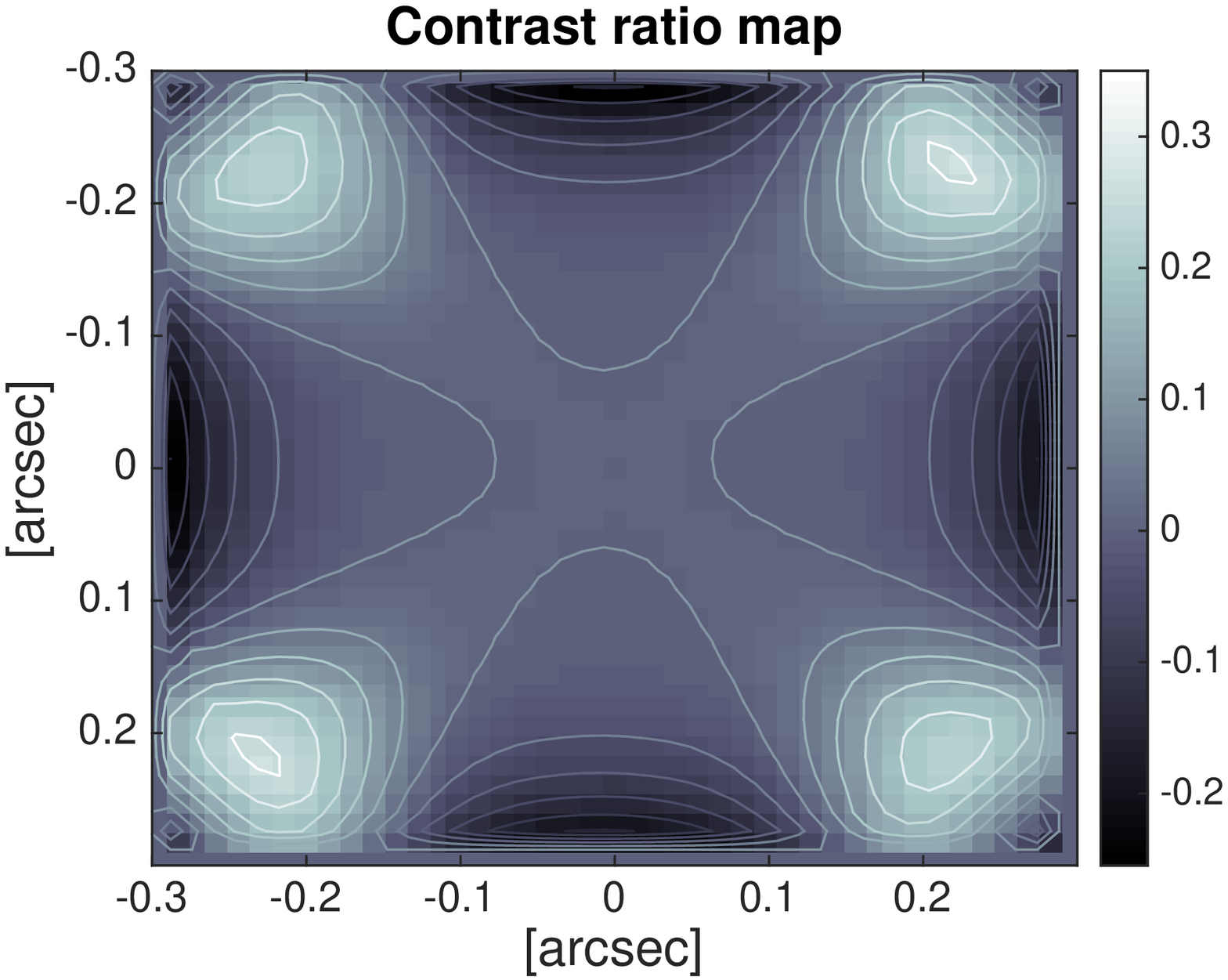}
         \end{tabular}
         \label{fig:contrast}
        \end{center}
	\caption[] 
	{\label{fig:contrast} Maps illustrating improvement in post-coronagraphic contrast (log scale) for different
          controller methods on the Keck--II system.  Left: comparing an integrator controller
	and least-squares DKF.  A value of 0 indicates equivalent performance.  Here the DKF
	provides better contrast over most of the correctable band (contrast ratio $< 0$).
	Right: comparison of two DKF methods, one using a least-squares approach, the other
	applying knowledge of aliasing errors using an anti-aliasing filter.  The anti-aliasing method
	performs better in a cross-shaped region ($[\pm(5\,-\,10) \frac{\lambda}{D},0]$ and $[0,\pm(5\,-\,10) \frac{\lambda}{D}]$)
	whilst the least-squares spatial filter coupled with the DKF
        performs slightly better in the corners.
     }
\end{figure*}

\section{Summary}\label{sec:summary}

We provide a general formulation for evaluating AO-related
wave-front residuals in the spatial-frequency domain. We start by outlining how the spatio-temporal
nature of the wave-front can be integrated on a controller for
optimised rejection, developing power-spectral
density formulae for four main error items: fitting, aliasing, noise and
aniso-servo-lag. We show how any of these three errors can be rejected
optimally using an anti-aliasing Wiener filter coupled to a predictive
closed-loop controller. We provide a suitable formulation for the
Distributed-Kalman-Filter (DKF) which is
synthesised in parallel, mode-by-mode, under a flexible and convenient state-space framework for
the stratified atmosphere case. 

Assuming perfect knowledge of the wind at all heights, the DKF   
with anti-aliasing Wiener filter allows for Strehl-optimal
performance, achieving the best trade-off between aliasing, noise
and aniso-servo-lag errors. Post-facto aliasing rejection may be
particularly useful when optical spatial filters are of little practical use and one relies therefore solely on
numerical processing of signals.

We show that contrast improvements of up to three orders of
magnitude can be achieved on 10\,m-class telescopes with a 0-magnitude
star at few $\lambda/D$ separation, typically $1-5 \lambda/D$. For the Keck-II AO system, we
show a wave-front error reduction of $\sim$60\,nm\,rms leading to
contrast enhancement of roughly one order of magnitude at separations
$1-5 \lambda/D$, by removing the "butterfly''-shaped servo-lag residual
commonly observed in high-contrast imaging. 

Finally, the developments herein can be straightforwardly used to
obtaining the real-time optimal controller and the power-spectral
densities used as filters when reconstructing the AO-corrected PSF
from system telemetry. 




\section{Software packages}

All the simulations and analysis were done with the object- oriented Matlab AO simulator (OOMAO) \cite{conanr14}. The class
  \verb+spatialFrequencyAdaptiveOptics+ implementing the analytics
  developed in this paper as well as the results herein is packed with
  the end-to-end library freely available from
  \htmladdnormallink{https://github.com/cmcorreia/LAM-Public}{https://github.com/cmcorreia/LAM-Public}. 

\section*{Acknowledgements}		

The research leading to these results 
received the support of the A*MIDEX
project (no. ANR-11-IDEX-0001- 02) funded by the "Investissements
d’Avenir'' French Government program, managed by the French National
Research Agency (ANR). 

\appendix

\section{LQG transfer functions}\label{sec:lqgTFs}
        \begin{example} \textbf{LQG transfer functions.}

          The state-update and state-estimate equations of the LQG are
          \begin{align}
          \label{eq:Kalman_estimation_update}
            \left\{\begin{array}{cl}
                \widehat{\xvec}_{k|k} & = \widehat{\xvec}_{k|k-1} + \Hkal (\svec_k - \Cd
                \widehat{\xvec}_{k|k-1} + \mathcal{D} u_{k-d} ) \\
                \widehat{\xvec}_{k+1|k} & = \Ad \widehat{\xvec}_{k|k} + \Bd \uvec_k
              \end{array}\right. ,
          \end{align}
          The second line of Eq.~(\ref{eq:Kalman_estimation_update})
          can be likewise written, using $\widehat{\xz}_p$ and
          $\widehat{\xz}_e$ the $\mathcal{Z}$-transforms of
          $\widehat{\xvec}_{k|k}$ and $\widehat{\xvec}_{k|k-1}$ respectively.
          \begin{equation}
            \z \widehat\xz_{p}  = \Ad \widehat{\xz}_{e} + \Bd \uz.
          \end{equation}
Let's start by assuming that the state cannot be attained by the
          controls, i.e. $\Bd = \0$. It is generalized afterwards.

          Multiplying out by $z^{-1}$ on both sides, one gets 
          \begin{equation}
            \widehat{\xz}_{p} = \z^{-1} \Ad \widehat{\xz}_{e}.
          \end{equation}
          With this result and the second line in Eq.~(\ref{eq:Kalman_estimation_update})
          \begin{align}
              \widehat{\xz}_{p} & 
              = \z^{-1} \Ad  \left(\widehat{\xz}_{p} +
              	\Hkal \left( \yz - \Cd \widehat{\xz}_{p}  + \z^{-d}
                  \mathcal{D} \uz\right)\right) \nonumber \\
              & = \left(\I -  \z^{-1}\Ad \left( \I - \Hkal \Cd
                \right)\right)^{-1} \z^{-1} \Ad \Hkal \nonumber \\&
                                                                    \qquad \times
                                                                                                                                        \left(
                                                                      \z^{-d}
                                                                      \mathcal{D}
                                                                      \uz
                                                                      +
                                                                      \yz\right)
                                                                    .
          \end{align}
          
          Using a command vector $\uvec_k = -\K \widehat{\xvec}_{k|k-1}$,
	i.e. a \textbf{causal form for the controls},
	one gets 
	\begin{equation}
  		\begin{split} \uz = & - \left[ \I + z^{-1}\K \Lambda_{p} \left(\z^{-d} \Lkal
        		\mathcal{D} + \Bd \right)\right]^{-1} 
                    \\ &\qquad \times
                    z^{-1} \K
            	\Lambda_{p} \Ad \Hkal \yz.
		\end{split}
	\end{equation}
	where 
  	\begin{equation}
    		\Lambda_{p} = \left(\I -  \z^{-1}\Ad \left( \I - \Hkal \Cd
              	\right)\right)^{-1} .
  	\end{equation}
	from which the controller transfer function writes
  	\begin{equation}
		\begin{split}
    			\mathbf{C}_p(\z) = &  - \left[ \I + z^{-1}\K \Lambda_{p} \left(\z^{-d} \Lkal
      			\mathcal{D} + \Bd \right)\right]^{-1}  \\
                    &\qquad \times
                    z^{-1}\K
            		\Lambda_{p} \Ad \Hkal  
		\end{split}.
  	\end{equation}

 	If instead we were to compute $\uvec_k = -\K \widehat{\xvec}_{k+1|k}$ then
          \begin{equation}
		\begin{split}
            		\uz = & -\K \left(\I - \Ad \left( \I - \Hkal \Cd
              		\right)\right)^{-1}  \\ &\qquad \times
                      \left( \z^{-d} \mathcal{D}  \uz + \Ad \Hkal \yz\right).
          	\end{split}
	\end{equation}
         Grouping terms, one gets
          \begin{equation}
            	\uz = - \left( \I + \K \Lambda_{p} \z^{-2} \Lkal \mathcal{D} \right)^{-1} \K
            	\Lambda_{p} \Ad \Hkal \yz,
          \end{equation}
  
 	The controller transfer function is finally written as
  	\begin{equation}
    		\label{eq:Kalman_predictor_TFdemo}
    		\mathbf{C}_p(\z) =  - \left( \I + \K \Lambda_{p} \z^{-d} \Lkal \mathcal{D} \right)^{-1} \K
            	\Lambda_{p} \Ad \Hkal  .
  	\end{equation}
	The general case in \textbf{predictor form} with an arbitrary delay and  $\Bd \neq \0$ one gets
  	\begin{equation}
		\begin{split}
    			\mathbf{C}_p(\z) = & - \left[ \I + \K \Lambda_{p} \left(\z^{-d} \Lkal
      			\mathcal{D} + \Bd \right)\right]^{-1}  \\
                    &\qquad \times
                    \K
            		\Lambda_{p} \Ad \Hkal  .
  		\end{split}
	\end{equation}

	For a command vector in \textbf{estimator form} $\uvec_k = -\K
  	\widehat{\xvec}_{k|k}$, following an analogous reasoning 
  	\begin{equation}
		\begin{split}
    			\widehat{\xz}_{e} & 
    			= \left(\I - \z^{-1}\Ad + \z^{-1} \Hkal \Cd
                          \Ad \right)^{-1}  \\ &\qquad \times
    			\Ad \Hkal \left( \z^{-d} \mathcal{D}  \uz +  \yz\right),
		\end{split}  
	\end{equation}       
  	yielding
  	\begin{equation}
		\begin{split}
    			\uz = & -\K \left(\I - \z^{-1}\Ad + \z^{-1} \Hkal \Cd \Ad
    			\right)^{-1}  \\ & \qquad \times
    			\Ad \Hkal \left( \z^{-d} \mathcal{D}  \uz +  \yz\right)
		\end{split}  
	\end{equation}
  	Grouping terms
 	\begin{equation}
    		\mathbf{C}_e(\z)  = - \left(\I + \K \Lambda_{e} \z^{-d}\Hkal \mathcal{D}\right)^{-1} \K \Lambda_{e} \Hkal,
  	\end{equation}
	with 
  	\begin{equation}
    		\Lambda_{e} = \left(\I - \z^{-1}\Ad + \z^{-1} \Hkal \Cd \Ad \right)^{-1}.
  	\end{equation}

	The general case when  $\Bd \neq \0$ and arbitrary delay equates to
 	\begin{equation}
		\begin{split}
    			\mathbf{C}_e(\z)  = & - \left\{\I + \K \Lambda_{e} \left[z^{-1}
        			(\I-\Hkal\Cd) \Bd \right.\right.+ \\ &\qquad
                                 \left.\left.\z^{-d}\Hkal\mathcal{D}\right]
    			\right\}^{-1}  
                        \K \Lambda_{e} \Hkal,
		\end{split}
  	\end{equation}
 	$\qed$

\end{example}


\begin{thebibliography}{0}%
\makeatletter
\providecommand \@ifxundefined [1]{%
 \@ifx{#1\undefined}
}%
\providecommand \@ifnum [1]{%
 \ifnum #1\expandafter \@firstoftwo
 \else \expandafter \@secondoftwo
 \fi
}%
\providecommand \@ifx [1]{%
 \ifx #1\expandafter \@firstoftwo
 \else \expandafter \@secondoftwo
 \fi
}%
\providecommand \natexlab [1]{#1}%
\providecommand \enquote  [1]{``#1''}%
\providecommand \bibnamefont  [1]{#1}%
\providecommand \bibfnamefont [1]{#1}%
\providecommand \citenamefont [1]{#1}%
\providecommand \href@noop [0]{\@secondoftwo}%
\providecommand \href [0]{\begingroup \@sanitize@url \@href}%
\providecommand \@href[1]{\@@startlink{#1}\@@href}%
\providecommand \@@href[1]{\endgroup#1\@@endlink}%
\providecommand \@sanitize@url [0]{\catcode `\\12\catcode `\$12\catcode
  `\&12\catcode `\#12\catcode `\^12\catcode `\_12\catcode `\%12\relax}%
\providecommand \@@startlink[1]{}%
\providecommand \@@endlink[0]{}%
\providecommand \url  [0]{\begingroup\@sanitize@url \@url }%
\providecommand \@url [1]{\endgroup\@href {#1}{\urlprefix }}%
\providecommand \urlprefix  [0]{URL }%
\providecommand \Eprint [0]{\href }%
\providecommand \doibase [0]{http://dx.doi.org/}%
\providecommand \selectlanguage [0]{\@gobble}%
\providecommand \bibinfo  [0]{\@secondoftwo}%
\providecommand \bibfield  [0]{\@secondoftwo}%
\providecommand \translation [1]{[#1]}%
\providecommand \BibitemOpen [0]{}%
\providecommand \bibitemStop [0]{}%
\providecommand \bibitemNoStop [0]{.\EOS\space}%
\providecommand \EOS [0]{\spacefactor3000\relax}%
\providecommand \BibitemShut  [1]{\csname bibitem#1\endcsname}%
\let\auto@bib@innerbib\@empty
\end{thebibliography}%


\begin{thebibliography}{10}
\newcommand{\enquote}[1]{``#1''}

\bibitem{correia14a}
C.~M. {Correia} and J.~{Teixeira}, \enquote{{Anti-aliasing Wiener filtering for
  wave-front reconstruction in the spatial-frequency domain for high-order
  astronomical adaptive-optics systems},} Journal of the Optical Society of
  America A \textbf{31}, 2763 (2014).

\bibitem{rigaut98}
F.~J. Rigaut, J.-P. V{\'e}ran, and O.~Lai, \enquote{{Analytical model for
  Shack-Hartmann-based adaptive optics systems},} in \enquote{Proc. of the
  SPIE,} , vol. 3353, D.~Bonaccini and R.~K. Tyson, eds. (SPIE, 1998), vol.
  3353, pp. 1038--1048.

\bibitem{flicker07a}
R.~Flicker, \enquote{Analytical evaluations of closed-loop adaptive optics
  spatial power spectral densities,} Tech. rep., W. M. Keck Observatory (2007).

\bibitem{jolissaint10}
L.~{Jolissaint}, \enquote{{Synthetic modeling of astronomical closed loop
  adaptive optics},} Journal of the European Optical Society - Rapid
  publications, 5, 10055 \textbf{5} (2010).

\bibitem{ellerbroek05}
B.~L. Ellerbroek, \enquote{Linear systems modeling of adaptive optics in the
  spatial-frequency domain,} J. Opt. Soc. Am. A \textbf{22}, 310--322 (2005).

\bibitem{rigaut00}
F.~J. Rigaut, B.~L. Ellerbroek, and R.~Flicker, \enquote{Principles,
  limitations, and performance of multiconjugate adaptive optics,} in
  \enquote{Proc. of the SPIE,} , vol. 4007 (2000), vol. 4007, pp. 1022--1031.

\bibitem{gavel04}
D.~T. {Gavel}, \enquote{{Tomography for multiconjugate adaptive optics systems
  using laser guide stars},} in \enquote{Proc. of the SPIE,} ,
vol. 5490, pp. 1356--1373 (2004).

\bibitem{ono16}
Y.~H. {Ono}, C.~M. {Correia}, D.~R. {Andersen}, O.~{Lardi{\`e}re}, S.~{Oya},
  M.~{Akiyama}, K.~{Jackson}, and C.~{Bradley}, \enquote{{Statistics of
  turbulence parameters at Maunakea using the multiple wavefront sensor data of
  RAVEN},} Monthly Notices of the Royal Astronomical Society,\textbf{465}, 4931--4941 (2017).

\bibitem{poyneer09}
L.~Poyneer, M.~van Dam, and J.-P. V{\'e}ran, \enquote{Experimental verification
  of the frozen flow atmospheric turbulence assumption with use of astronomical
  adaptive optics telemetry,} J. Opt. Soc. Am. A \textbf{26}, 833--846 (2009).

\bibitem{poyneer07}
L.~A. Poyneer, B.~A. Macintosh, and J.-P. V{\'e}ran, \enquote{{Fourier
  transform wavefront control with adaptive prediction of the atmosphere},} J.
  Opt. Soc. Am. A \textbf{24}, 2645--2660 (2007).

\bibitem{massioni11}
P.~Massioni, C.~Kulcs\'{a}r, H.-F. Raynaud, and J.-M. Conan, \enquote{Fast
  computation of an optimal controller for large-scale adaptive optics,} J.
  Opt. Soc. Am. A \textbf{28}, 2298--2309 (2011).

\bibitem{oppenheim97}
A.~V. Oppenheim and A.~S. Willsky, \emph{Signals \& Systems}
  (Prentice-Hall,Inc., 1997), 2nd ed.

\bibitem{oppenheim99}
A.~V. {Oppenheim} and R.~W. {Schafer}, \emph{Discrete-time signal processing}
  (Prentice-Hall,Inc., 1999), 2nd ed.

\bibitem{poyneer04a}
L.~A. Poyneer and B.~Macintosh, \enquote{Spatially filtered wave-front sensor
  for high-order adaptive optics,} J. Opt. Soc. Am. A \textbf{21}, 810--819
  (2004).

\bibitem{sauvage16}
J.-F. {Sauvage}, T.~{Fusco}, C.~{Petit}, A.~{Costille}, D.~{Mouillet}, J.-L.
  {Beuzit}, K.~{Dohlen}, M.~{Kasper}, M.~{Suarez}, C.~{Soenke}, A.~{Baruffolo},
  B.~{Salasnich}, S.~{Rochat}, E.~{Fedrigo}, P.~{Baudoz}, E.~{Hugot},
  A.~{Sevin}, D.~{Perret}, F.~{Wildi}, M.~{Downing}, P.~{Feautrier},
  P.~{Puget}, A.~{Vigan}, J.~{O'Neal}, J.~{Girard}, D.~{Mawet}, H.~M. {Schmid},
  and R.~{Roelfsema}, \enquote{{SAXO: the extreme adaptive optics system of
  SPHERE (I) system overview and global laboratory performance},} Journal of
  Astronomical Telescopes, Instruments, and Systems \textbf{2}, 025003 (2016).

\bibitem{guyon05}
O.~Guyon, \enquote{Limits of adaptive optics for high-contrast imaging,} The
  Astrophysical Journal \textbf{629}, 592 (2005).

\bibitem{poyneer08a}
L.~A. Poyneer and J.-P. V{\'e}ran, \enquote{Toward feasible and effective
  predictive wavefront control for adaptive optics,} in \enquote{Proc. of the
  SPIE - Adaptive Optical Systems,} , vol. 7015, N.~Hubin, C.~E. Max, and P.~L.
  Wizinowich, eds. (SPIE, 2008), vol. 7015, p. 70151E.

\bibitem{madec99}
P.-Y. Madec, \emph{{Adaptive Optics for Astronomy}} (Cambridge University
  Press, New York, 1999), chap. Control Techniques.

\bibitem{clare06}
R.~M. Clare, B.~L. Ellerbroek, G.~Herriot, and J.-P. V\'{e}ran,
  \enquote{Adaptive optics sky coverage modeling for extremely large
  telescopes,} Appl. Opt. \textbf{45}, 8964--8978 (2006).

\bibitem{roddier99}
F.~Roddier, \emph{Adaptive Optics in Astronomy} (Cambridge University Press,
  New York, 1999).

\bibitem{correia10a}
C.~Correia, H.-F. Raynaud, C.~Kulcs{\'a}r, and J.-M. Conan, \enquote{On the
  optimal reconstruction and control of adaptive optical systems with mirroir
  dynamics,} J. Opt. Soc. Am. A \textbf{27}, 333--349 (2010).

\bibitem{poyneer05}
L.~A. Poyneer and J.-P. V{\'e}ran, \enquote{{Optimal modal Fourier-transform
  wavefront control},} J. Opt. Soc. Am. A \textbf{22}, 1515--1526 (2005).

\bibitem{poyneer10}
L.~A. Poyneer and J.-P. V\'{e}ran, \enquote{Kalman filtering to suppress
  spurious signals in adaptive optics control,} J. Opt. Soc. Am. A \textbf{27},
  A223--A234 (2010).

\bibitem{kulcsar09}
C.~Kulcs{\'a}r, H.-F. Raynaud, J.-M. Conan, C.~Correia, and C.~Petit,
  \enquote{Control design and turbulent phase models in adaptive optics: A
  state-space interpretation,} in \enquote{Adaptive Optics: Methods, Analysis
  and Applications,}  (Optical Society of America, 2009), p. AOWB1.

\bibitem{hardy98}
J.W.Hardy, \emph{Adaptive Optics for Astronomical Telescopes} (Oxford, New
  York, 1998).

\bibitem{thomas06}
S.~{Thomas}, T.~{Fusco}, A.~{Tokovinin}, M.~{Nicolle}, V.~{Michau}, and
  G.~{Rousset}, \enquote{{Comparison of centroid computation algorithms in a
  Shack-Hartmann sensor},} Monthly Notices of the Royal Astronomical Society, \textbf{371}, 323--336 (2006).

\bibitem{perrin03}
M.~D. Perrin, A.~Sivaramakrishnan, R.~B. Makidon, B.~R. Oppenheimer, and J.~R.
  Graham, \enquote{The structure of high strehl ratio point-spread functions,}
  The Astrophysical Journal \textbf{596}, 702 (2003).

\bibitem{macintosh05}
B.~Macintosh, L.~Poyneer, A.~Sivaramakrishnan, and C.~Marois, \enquote{Speckle
  lifetimes in high-contrast adaptive optics,}  (2005).

\bibitem{herscovici17}
O.~{Herscovici-Schiller}, L.~M. {Mugnier}, and J.-F. {Sauvage}, \enquote{{An
  analytic expression for coronagraphic imaging through turbulence. Application
  to on-sky coronagraphic phase diversity},} Monthly Notices of the Royal Astronomical Society, \textbf{467}, L105--L109
  (2017).

\bibitem{conanr14}
R.~{Conan} and C.~{Correia}, \enquote{{Object-oriented Matlab adaptive optics
  toolbox},} in \enquote{Society of Photo-Optical Instrumentation Engineers
  (SPIE) Conference Series,} , vol. 9148, p.~6 (2014).

\end{thebibliography}
\end{document}